\documentclass[12pt,a4paper]{article}

\usepackage[utf8]{inputenc}
\usepackage[T1]{fontenc}
\usepackage{amsmath,amssymb,amsfonts,mathtools}
\usepackage[margin=1in]{geometry}
\usepackage{booktabs}
\usepackage{hyperref}
\usepackage{natbib}
\usepackage{setspace}
\usepackage{graphicx}
\usepackage{float}
\usepackage{caption}
\usepackage{array}
\usepackage{tabularx}
\usepackage{doi}
\usepackage{url}
\usepackage{listings}
\usepackage{xcolor}
\usepackage{fancyhdr}
\usepackage{graphicx}

\hypersetup{
    colorlinks=true,
    linkcolor=blue!70!black,
    citecolor=blue!70!black,
    urlcolor=blue!70!black,
    pdftitle={Cheaper AI, More Informality?},
    pdfauthor={Gabriel Montes-Rojas, Fernando Toledo}
}

\lstdefinestyle{dynare}{
    language=Matlab,
    basicstyle=\ttfamily\scriptsize,
    keywordstyle=\color{blue!70!black}\bfseries,
    commentstyle=\color{green!50!black}\itshape,
    stringstyle=\color{red!70!black},
    numbers=left,
    numberstyle=\tiny\color{gray},
    numbersep=5pt,
    frame=single,
    breaklines=true,
    breakatwhitespace=false,
    tabsize=4,
    showstringspaces=false,
    columns=flexible,
    xleftmargin=2em,
    framexleftmargin=1.5em,
    aboveskip=1em,
    belowskip=1em,
    morekeywords={var,varexo,parameters,model,end,initval,steady,check,shocks,
                  stoch_simul,stderr,long_name}
}

\title{\textbf{Cheaper AI, More Informality?\\ A Dual Labor Market Model for Developing Economies}}
\author{
    Gabriel Montes-Rojas\thanks{Instituto Interdisciplinario de Econom\'{i}a Pol\'{i}tica de Buenos Aires (IIEP--BAIRES), Facultad de Ciencias Econ\'{o}micas, Universidad de Buenos Aires and CONICET. Email: \texttt{gabriel.montes@economicas.uba.ar}}
    \and
    Fernando Toledo\thanks{Corresponding author. Facultad de Ciencias Econ\'{o}micas, Universidad Nacional de La Plata (FCE--UNLP). Email: \texttt{fernando.toledo@econo.unlp.edu.ar}.}
    \and
    Juan Manuel Rodriguez Repeti\thanks{Instituto Interdisciplinario de Econom\'{i}a Pol\'{i}tica de Buenos Aires (IIEP--BAIRES), Facultad de Ciencias Econ\'{o}micas, Universidad de Buenos Aires and CONICET. Email: \texttt{rodriguezrepeti.jm@economicas.uba.ar}}
}
\date{May 2026}

\begin{document}

\maketitle

\begin{abstract}
This paper studies what happens when AI gets cheaper, with emphasis on the labor market outcomes, whether it creates formal jobs or whether it pushes workers into informality. We argue that the answer depends on the elasticity of substitution between imported AI capital and formal labor. We build a small open economy DSGE model with a dual labor market, imported AI capital, and country risk, calibrated to an economy where informality is pervasive. The same decline in AI prices produces sharply different labor-market outcomes depending on whether AI substitutes or complements formal workers. Under substitution, cheaper AI weakens formal labor demand and increases the role of the informal sector as an employment buffer. Under complementarity, it expands formal employment and amplifies output, wages, investment, and capital accumulation. The model therefore shows that AI can become either a source of displacement pressure or a driver of formal-sector expansion, depending on how it interacts with human labor.


\end{abstract}

\newpage

%% ============================================================
\section{Introduction}
%% ============================================================

What actually happens in the labor market when artificial intelligence (AI) gets cheaper and more accessible? Does it create good, formal jobs, or does it simply push more people into the informal sector?

This question carries enormous implications for inequality, productivity, and social stability. In countries where informal work makes up over 40 percent of the labor force---and in much of Latin America it approaches 50 percent or more---the answer matters deeply. In Bolivia, Peru, or Honduras, more than two-thirds of workers operate informally \citep{ILO2024}. If AI displaces formal workers, it could swell the ranks of the informal sector, eroding the tax base and leaving more people without social protections. If AI complements formal labor, cheaper technology could actually accelerate formalization and growth.

Most standard macroeconomic models cannot adequately address this question. These models, built for developed economies with unified labor markets, fail to account for the dual labor market structure that defines much of Africa, Asia and Latin America, where formal workers have legal protections and pay taxes while informal workers operate outside the formal economic system. And the newer DSGE models that examine AI tend to assume fixed factor shares, so they cannot capture whether AI substitutes for or complements human labor.

As \citet{AcemogluRestrepo2018} emphasize, everything comes down to a single deep parameter: the elasticity of substitution between AI capital and labor. Get that parameter wrong, and the policy response will be exactly backwards.

In this paper, we develop a small open economy DSGE model that brings together three features essential for understanding developing economies: a dual labor market where workers can move between formal and informal jobs, imported AI capital used only in the formal sector, and a country risk premium tied to external debt. We calibrate the model to Latin American data, drawing on the automation literature while adapting it to the structural realities of a region where informality is pervasive.

The substitution elasticity serves as the pivotal parameter for the labor-market transmission of an AI price shock. Under complementarity, cheaper AI boosts demand for both AI inputs and formal labor, generating stronger increases in output, wages, investment, and capital accumulation. Under substitution, the direct displacement effect weakens formal labor demand and increases the role of the informal sector as an employment buffer. However, the aggregate response is shaped by general equilibrium forces: relative prices, demand spillovers, capital accumulation, and the external debt channel can amplify or dampen the initial labor-market effect. The magnitude of these effects depends on factors such as labor mobility and the nature of informal production, but the basic direction remains clear. 

We simulate a 1 percent drop in the international price of AI under both regimes, using quarterly data and standard parameter values from the literature. The results bear out the theory. They show that the informal sector acts as a buffer whose response depends on both technology and demand-side forces, it absorbs displaced workers when AI substitutes for labor, while under complementarity its dynamics are shaped by the expansion of aggregate income and demand for non-tradable goods. This buffering role has important aggregate welfare implications, although the representative-agent structure prevents us from drawing distributional conclusions across worker types. The model also shows that complementarity and substitution in consumption goods is also important in studying the effects of AI.

This paper makes three main contributions. First, we embed the task-based insights of \citet{AcemogluRestrepo2018} into a DSGE model with dual labor markets, providing a framework to analyze how AI adoption affects informality---a critical issue for Latin America. Second, we offer a rigorous calibration for a representative Latin American economy, drawing on recent data from the region, and we provide fully replicable Dynare code so other researchers can adapt it to their own contexts. Third, we derive analytical steady-state conditions that link informal sector parameters to observable targets, which should facilitate future empirical work.

Our contribution is not to show that informality buffers labor-market shocks, a mechanism already established in dual-sector small open economy DSGE models \citep[e.g.,][]{FernandezMeza2015, HorvathYang2022}. Instead, we identify a new transmission channel: AI is modeled as an imported intermediate capital input, so international AI price shocks affect labor reallocation through its elasticity of substitution with formal labor. This channel determines when the informal-sector buffer is activated, a mechanism absent from prior dual-labor-market DSGE models.

The paper proceeds as follows. Section~\ref{sec:facts} presents stylized facts about AI and labor markets in Latin America. Section~\ref{sec:literature} reviews the relevant literature and positions our contribution. Section~\ref{sec:model} lays out the model. Section~\ref{sec:equilibrium} defines the competitive equilibrium and summarizes the main transmission mechanism. Section~\ref{sec:empirical} reviews related empirical evidence. Section~\ref{sec:calibration} describes the calibration strategy and solution method. Section~\ref{sec:results} presents the simulation results. Section~\ref{sec:policy} discusses policy implications. Section~\ref{sec:conclusion} concludes.

%% ============================================================
\section{Stylized Facts: AI and Labor Markets in Latin America}\label{sec:facts}
%% ============================================================

We focus on Latin America because this is a region where the dual labor market structure is the rule. 

Below we comment on some stylized facts that are expected features of our modeling choices. 
These 4 facts motivate our modeling choices: a dual labor market (Fact 1), the substitution elasticity as pivotal parameter (Fact 2), explicit labor mobility (Fact 3), and distributional concerns (Fact 4).

\vspace{.5cm}

\noindent\textit{Fact 1: Informality dominates Latin America, and this dominance shapes how technology gets adopted.}

Latin America and the Caribbean has some of the highest informality rates in the world, and it is also one of the most unequal regions \citep{GaspariniTornarolli2009}. In Bolivia, Peru, and Honduras, more than two-thirds of workers are informal \citep{ILO2024}. This means that most people work without social protection, without labor benefits, and outside the formal financial system---precisely the kinds of institutions that help workers adapt when technology shifts beneath their feet. Informal businesses are typically small, lack access to credit, and operate in the shadows of the law, making it difficult for them to invest in new technologies \citep{Ulyssea2018}.

\vspace{.5cm}

\noindent\textit{Fact 2: Roughly a quarter to a third of Latin American jobs could feel the effects of Generative AI, but full automation remains unlikely for most.}

The ILO and World Bank \citep{ILO2024} find that between 26 and 38 percent of jobs in the region could experience influence from Generative AI. The crucial nuance: only 2 to 5 percent of jobs face serious automation risk, while 8 to 14 percent could see productivity improvements. \citet{Azuara2024} examined Chile, Mexico, and Peru using GPT-4-based exposure measures, finding that about three-quarters of jobs have at least some exposure ($\geq$10\% of tasks), but substantial exposure ($\geq$40\% of tasks) drops to 6--20 percent. \citet{Ciascchi2025} expanded the analysis to 14 countries, confirming that AI exposure concentrates in specific occupations and sectors.

\vspace{.5cm}

\noindent\textit{Fact 3: Informal workers face less exposure to AI, which protects them from displacement but also locks them out of productivity gains.}

A finding appears in every study we review: informality dramatically reduces measured AI exposure. Formal sector workers face much higher potential exposure---both to automation and to augmentation \citep{ILO2024}. The jobs most likely to benefit from AI augmentation fall in sales, architecture, education, health, and personal services---disproportionately formal occupations \citep{Azuara2024}.

\vspace{.5cm}

\noindent\textit{Fact 4: AI exposure does not distribute randomly. Women, younger workers, educated workers, and formal workers face higher exposure.}

Across 14 countries, \citet{Ciascchi2025} find that women, younger workers, educated workers, and formal workers consistently show higher AI exposure. \citet{Azuara2024} confirm these patterns in their three-country study. Table~\ref{tab:exposure} summarizes the key estimates.

\begin{table}[H]
\centering
\caption{Summary of AI Exposure Estimates for Latin America}\label{tab:exposure}
\small
\begin{tabularx}{\textwidth}{@{}l l X@{}}
\toprule
\textbf{Study} & \textbf{Countries} & \textbf{Key Findings} \\
\midrule
ILO--World Bank (2024) & Regional & 26--38\% of jobs exposed; 2--5\% automation risk; 8--14\% could see productivity gains; formal workers higher exposure. \\
\addlinespace
Azuara et al.\ (2024) & CL, MX, PE & 74--76\% jobs have some exposure ($\geq$10\% tasks); 6--20\% substantial ($\geq$40\% tasks); higher for women, educated, formal. \\
\addlinespace
Ciaschi et al. (2025) & 14 countries & Four AI indices; higher exposure for women, educated, formal workers; gradients flatten with complementarity adjustment. \\
\bottomrule
\end{tabularx}
\begin{flushleft}
\footnotesize\textit{Sources:} \citet{ILO2024}, \citet{Azuara2024}, \citet{Ciascchi2025}.
\end{flushleft}
\end{table}

%% ============================================================
\section{Literature Review}\label{sec:literature}
%% ============================================================

This paper draws on four distinct relatively new literatures. We briefly review each in turn, emphasizing how existing work motivates our modeling choices and where our contribution lies relative to prior efforts.

\subsection{Artificial Intelligence and Labor Markets}

The foundational work of \citet{AcemogluRestrepo2018} develops a task-based framework that distinguishes between two countervailing forces of automation: a \textit{displacement effect}, which reduces labor demand as machines take over tasks previously performed by workers, and a \textit{productivity effect}, which increases labor demand by raising output and creating new tasks where labor retains a comparative advantage. The elasticity of substitution between capital and labor emerges as the key parameter governing which force dominates. Subsequent empirical work by \citet{AcemogluRestrepo2020} provides the first causal estimates of robot adoption on local labor markets in the United States, finding that one additional robot per thousand workers reduces the employment-to-population ratio by 0.2 percentage points and wages by 0.42 percent---magnitudes that are consistent with the displacement effect dominating in robot-intensive sectors.

On the micro-empirical side, \citet{Autor2024} construct a novel database spanning eight decades of new job titles linked to U.S.\ Census microdata, showing that the majority of current employment is in occupations that did not exist in 1940 and that labor-augmenting innovations create substantial new work. \citet{BrynjolfssonLiRaymond2025} provide the first large-scale field evidence on generative AI in the workplace, finding that access to an AI conversational assistant raises worker productivity by 15 percent on average, with disproportionate gains accruing to less experienced and lower-skilled workers. This finding is directly relevant to our framework: if AI augments less-skilled workers, the effective $\sigma_{KL}$ may be closer to complementarity than to substitution, at least within certain occupations.

At the macro level, \citet{KarabarbounisNeiman2014} document the global decline of the labor share since the early 1980s, consistent with rising capital-labor substitutability in aggregate production. \citet{Felten2024} develop the AI Occupational Exposure Index (AIOE), linking AI capabilities to occupation-level abilities, and show that highly-educated, highly-paid, white-collar occupations face the greatest exposure---a pattern confirmed in developing-country settings by the Latin American studies we review in Section~\ref{sec:facts}. These findings underscore the need for models that accommodate regime-switching behavior, which is precisely what our framework captures through the pivotal parameter $\sigma_{KL}$.

\subsection{DSGE Models with AI}

The integration of AI into DSGE frameworks is still in its early stages. \citet{Berg2024} develop a three-sector model (advanced economies, emerging markets, and low-income countries) where AI impacts depend on sectoral exposure and data access. Their Cobb-Douglas production specification, however, imposes unit elasticity between capital and labor and therefore cannot generate the qualitatively different regimes that our CES nesting delivers. \citet{HemousOlsen2022} build an endogenous growth model where automation and horizontal innovation coexist, showing that the share of automating innovations increases endogenously over time as low-skill wages rise---a dynamic that our framework captures in reduced form through the exogenous $\sigma_{KL}$. At the global policy level, \citet{Cazzaniga2024} estimate that about 40 percent of employment worldwide is exposed to AI, with advanced economies facing both greater risks and greater potential benefits, while emerging markets may be partially shielded by lower cognitive-task intensity---a finding that resonates with the buffering role of informality in our model.

Our model differs from these contributions in three respects: (i)~it introduces dual labor markets with endogenous informality, (ii)~it treats AI as imported capital subject to international price shocks, and (iii)~it is calibrated to a representative Latin American economy with approximately 50 percent informality.

\subsection{Dual Labor Markets in Emerging Economies}

The literature on informality in developing countries has made substantial progress in recent years. \citet{Ulyssea2018} develops and estimates an equilibrium model with heterogeneous firms that choose whether to register their business (the extensive margin) and how many workers to hire off the books (the intensive margin), using matched employer-employee data from Brazil. His finding that informal firms are fundamentally different from formal ones---not merely tax-evading versions of the same entity---provides the microeconomic foundation for our simplifying assumption that the informal sector uses a distinct (labor-only) production technology. \citet{HaanwinckelSoares2021} extend this literature with a search model featuring heterogeneous workers and firms, labor regulations including minimum wage, and imperfect substitutability between skilled and unskilled labor. They estimate the model on Brazilian data and show that changes in workforce composition---particularly rising educational attainment---were the main driver of the decline in informality between 2003 and 2012. \citet{LaPortaShleifer2014} argue more broadly that informal firms in developing countries are genuinely low-productivity operations, not constrained entrepreneurs waiting to formalize.

Our model builds on this literature by endogenizing labor allocation between formal and informal sectors in response to \textit{technology shocks}---specifically, changes in the price of AI capital. This differs from the existing literature, which focuses on regulatory shocks (changes in enforcement, minimum wages, or payroll taxes) as the primary drivers of informality dynamics. \citet{GaspariniTornarolli2009} provide the empirical benchmarks for informality rates in the region that anchor our calibration.

\subsection{Small Open Economy Models}

We adopt the canonical debt-elastic interest rate specification of \citet{SchmittGroheUribe2003}, which ensures stationarity of external debt in small open economy models. The country risk premium follows \citet{NeumeyerPerri2005}, who show that interest rate shocks coupled with financial frictions are a primary source of business cycle fluctuations in emerging economies. \citet{AguiarGopinath2007} demonstrate that trend productivity shocks---rather than transitory fluctuations---are the dominant source of macroeconomic volatility in emerging markets, motivating our persistent AR(1) specification for both the productivity and AI price shocks. \citet{GarciaCicco2010} revisit this finding with over a century of Argentine and Mexican data and argue that financial frictions and country spreads play a larger role than previously recognized. Our model incorporates the debt-elastic spread but abstracts from financial frictions to maintain analytical tractability---a choice we revisit in the caveats.

Small open economy DSGE models in which a competitive informal sector absorbs workers as a buffer are already well established. Most directly, \citet{FernandezMeza2015} build a two-sector small open economy business-cycle model calibrated to Mexico, where the informal non-tradable, self-employed sector cushions formal-sector fluctuations. Related contributions include \citet{HorvathYang2022}, who show that informality dampens the response of unemployment to productivity and interest-rate shocks. What is new here is the transmission channel: AI enters as an imported intermediate capital input whose international price shock is filtered through the inner-CES substitution elasticity, thereby endogenously activating the informal-sector buffer. To our knowledge, no prior dual-labor-market DSGE model treats AI as a priced imported factor whose substitutability with formal labor governs the direction of labor reallocation.

%% ============================================================
\section{The Model}\label{sec:model}
%% ============================================================

We consider a small open economy populated by a representative household that supplies labor to two sectors---formal and informal---and consumes a composite good. The formal sector produces tradable goods using physical capital, formal labor, and imported AI capital. The informal sector produces non-tradable goods using only informal labor with decreasing returns to scale. The economy borrows from international capital markets at an interest rate that incorporates a country-specific risk premium.

This structure captures three features that are essential for analyzing AI adoption in Latin America. First, the dual labor market---with free mobility between formal and informal employment---allows us to trace how a technology shock in the formal sector propagates to the informal sector via labor reallocation. Second, the treatment of AI as imported capital subject to international price shocks reflects the reality that Latin American firms are predominantly technology \textit{importers}, not producers \citep[see the illustrative evidence in Section~\ref{sec:facts} and][]{AcemogluRestrepo2018}. Third, the debt-elastic interest rate captures the external vulnerability of emerging economies to capital flow volatility, a feature that amplifies the transmission of technology shocks in ways that closed-economy models cannot capture \citep{SchmittGroheUribe2003, NeumeyerPerri2005}.

\subsection{Household}\label{sec:household}

The representative household maximizes expected lifetime utility over an infinite horizon:
\begin{equation}
\mathbb{E}_0 \sum_{t=0}^{\infty} \beta^t \frac{\left(C_t - \chi \frac{L_t^{1+\nu}}{1+\nu}\right)^{1-\sigma}-1}{1-\sigma}
\label{eq:utility}
\end{equation}
where $C_t$ denotes aggregate consumption, $L_t$ is total hours worked, $\beta \in (0,1)$ is the subjective discount factor, $\sigma > 0$ governs the degree of relative risk aversion, $\nu > 0$ is the inverse of the Frisch elasticity of labor supply, and $\chi > 0$ is a scaling parameter for the disutility of labor. Following \citet{Greenwood1988} (GHH), this preference specification internalizes the labor supply decision within the consumption aggregator. Consequently, it neutralizes the wealth effect on labor supply, a feature that is essential for emerging economies experiencing large technology or terms-of-trade shocks, preventing positive innovations from generating a counterfactual contraction in aggregate labor hours.

Total labor supplied by the household is the sum of hours allocated to the formal and informal sectors:
\begin{equation}
L_t = L_{f,t} + L_{i,t}
\label{eq:labor_total}
\end{equation}

This specification implies that formal and informal hours are \textit{perfect substitutes in disutility}: the household cares about total hours worked, not about the sectoral composition of its labor supply. This is a deliberate simplification. It ensures that in equilibrium, the household is indifferent at the margin between supplying an additional hour to either sector, which delivers wage equalization across sectors---the key labor mobility condition that governs the formal-informal margin.

\paragraph{Consumption aggregation.} The household consumes a CES (constant elasticity of substitution) composite of tradable goods $C_{f,t}$ produced by the formal sector and non-tradable goods $C_{i,t}$ produced by the informal sector:
\begin{equation}
C_t = \left[ \omega^{\frac{1}{\eta}} C_{f,t}^{\frac{\eta-1}{\eta}} + (1-\omega)^{\frac{1}{\eta}} C_{i,t}^{\frac{\eta-1}{\eta}} \right]^{\frac{\eta}{\eta-1}}
\label{eq:CES_C}
\end{equation}
where $\omega \in (0,1)$ is the expenditure weight on tradable goods and $\eta > 0$ is the intratemporal elasticity of substitution between tradable and non-tradable goods. When $\eta = 1$, the aggregator reduces to Cobb-Douglas; when $\eta < 1$, the two goods are complements---meaning that the household has limited ability to substitute between formal and informal consumption when relative prices change.

Cost minimization by the household implies that the relative price of informal goods (which functions as the real exchange rate in this economy) equals\footnote{See Appendix A for the full derivation of this equation.}:
\begin{equation}
p_{i,t}=\left(\frac{1-\omega}{\omega}\right)^{\frac{1}{\eta}}\left(\frac{C_{f,t}}{C_{i,t}}\right)^{\frac{1}{\eta}}
\label{eq:relprice}
\end{equation}

This expression captures a key transmission channel: when the informal sector contracts (say, because workers move to formal employment), $C_{i,t}$ falls, driving up the relative price of informal goods $p_{i,t}$. This real exchange rate adjustment feeds back into the informal sector wage condition and into the household's consumption allocation, creating an endogenous stabilizing mechanism.

\paragraph{Consumer price index} The consumer price index (CPI), $P_t$, represents the minimum cost of purchasing one unit of the aggregate consumption bundle $C_t$. Given the CES aggregator \eqref{eq:CES_C} and taking the tradable good as the num\'{e}raire ($p_{f,t}=1$), the CPI is defined as:
\begin{equation}
P_t = \left[ \omega + (1-\omega) p_{i,t}^{1-\eta} \right]^{\frac{1}{1-\eta}}
\label{eq:CPI}
\end{equation}
Consequently, the household's total consumption expenditure is $P_t C_t = C_{f,t} + p_{i,t} C_{i,t}$.

\paragraph{Budget constraint.} The household's flow budget constraint, expressed in units of the tradable good (the num\'{e}raire), is:
\begin{equation}
P_t C_{t}+I_{t}+\frac{D_{t+1}}{1+r_{t}}+\frac{\phi}{2}(K_{t+1}-K_t)^2 = W_tL_t+R_t^kK_t+D_t+\Pi_{f,t}+\Pi_{i,t}
\label{eq:budget}
\end{equation}

The left-hand side records all uses of income: aggregate consumption $C_t$; physical investment $I_t$; purchases of one-period foreign bonds $D_{t+1}/(1+r_t)$ (where $D_{t+1}$ is the face value of debt carried into $t+1$ and $r_t$ is the interest rate determined in international markets); a quadratic capital adjustment cost $(\phi/2)(K_{t+1} - K_t)^2$ with parameter $\phi > 0$ following \citet{Christiano2005}.

The right-hand side records all sources of income: labor earnings at wage $W_t$ from both sectors (the wage is equalized by the free mobility condition); rental income from physical capital $R_t^k K_t$ (capital is predetermined, so $K_t$ is available for production at time $t$); maturing foreign bonds $D_t$; and profits from both formal ($\Pi_{f,t}$) and informal firms ($\Pi_{i,t}$).

Physical capital accumulates according to the standard law of motion:
\begin{equation}
K_{t+1}=(1-\delta)K_t+I_t.
\label{eq:capital_accum}
\end{equation}
where $\delta \in (0,1)$ is the depreciation rate.

\subsection{Household Optimization Problem}

The representative household chooses the paths of $\{C_t, L_t, I_t, K_t, D_{t+1}\}_{t=0}^{\infty}$ to maximize its expected lifetime utility (Equation \ref{eq:utility}) subject to No-Ponzi condition and the budget constraints:
\begin{equation}
P_tC_t+K_{t+1}-(1-\delta)K_t+\frac{D_{t+1}}{1+r_t}
+\frac{\phi}{2}(K_{t+1}-K_t)^2 = W_tL_t+R_t^kK_t+D_t+\Pi_{f,t}+\Pi_{i,t}
\end{equation}
where we have substituted $I_t=K_{t+1}-(1-\delta)K_t$ into the original budget constraint.

\subsubsection{The Lagrangian}

Let $\beta^t \lambda_t$ be the Lagrange multiplier associated with the budget constraint at time $t$. The Lagrangian of the household's problem is defined as:

\begin{align}
\mathcal{L} = \mathbb{E}_0 \sum_{t=0}^\infty \beta^t \Biggl\{
& \frac{\left(C_t - \chi \frac{L_t^{1+\nu}}{1+\nu}\right)^{1-\sigma}-1}{1-\sigma} \notag \\
& +\lambda_t\left[
W_tL_t+R_t^kK_t+D_t+\Pi_{f,t}+ \right. \notag \\ 
&\left. \Pi_{i,t}
-P_tC_t -\left(K_{t+1}-(1-\delta)K_t\right)
-\frac{D_{t+1}}{1+r_t}
-\frac{\phi}{2}(K_{t+1}-K_t)^2 \right]
\Biggr\}
\end{align}

\subsubsection{First-Order Conditions (FOCs)}

The first-order conditions with respect to $C_t$, $L_t$, $D_{t+1}$, and $K_t$ are given by:

\begin{enumerate}
    \item \textbf{Consumption ($C_t$):}
    \begin{equation}
    \frac{\partial \mathcal{L}}{\partial C_t} = 0 \implies \lambda_t = \frac{\left(C_t - \chi \frac{L_t^{1+\nu}}{1+\nu}\right)^{-\sigma}}{P_t}
    \label{eq:FOC_C}
    \end{equation}
Where $\lambda_t$ is the Lagrange multiplier that represents the marginal utility of an additional unit of tradable income, which depends on both the level of consumption and the current price of the aggregate bundle.

    \item \textbf{Labor Supply ($L_t$):}
    \begin{equation}
    \frac{\partial \mathcal{L}}{\partial L_t} = 0 \implies \chi L_t^\nu \left(C_t - \chi \frac{L_t^{1+\nu}}{1+\nu}\right)^{-\sigma} = \lambda_t W_t
    \label{FOC_L_households}
    \end{equation}
This equates the marginal disutility of labor (left-hand side) to the marginal value of labor income (right-hand side). The parameter $\nu$ governs the curvature of labor disutility: a higher $\nu$ implies that labor supply responds less elastically to wage changes.

    \item \textbf{External Debt ($D_{t+1}$):}
    \begin{equation}
    \frac{\partial \mathcal{L}}{\partial D_{t+1}} = 0 \implies \lambda_t = \beta (1+r_t) \mathbb{E}_t [ \lambda_{t+1} ]
    \label{eq:FOC_D}
    \end{equation}
This is the standard consumption Euler equation for a small open economy. It equates the marginal utility of consuming one unit today ($\lambda_t$) to the expected discounted marginal utility of saving that unit and consuming $(1+r_t)$ units tomorrow. In steady state, this implies $r = 1/\beta - 1$, which pins down the steady-state interest rate.

    \item \textbf{Physical Capital ($K_{t+1}$):}
    \begin{equation}
    \frac{\partial \mathcal{L}}{\partial K_{t+1}} = 0 \implies \lambda_t\left[1+\phi(K_{t+1}-K_t)\right] = \beta \mathbb{E}_t\left[\lambda_{t+1}\left(R_{t+1}^k+1-\delta+\phi(K_{t+2}-K_{t+1})\right)\right]
    \label{eq:FOC_K}
    \end{equation}
\end{enumerate}

The left-hand side is the marginal cost of investment: one unit of forgone consumption plus the marginal adjustment cost $\phi(K_{t+1} - K_t)$. The right-hand side is the expected marginal benefit: the rental rate $R_{t+1}^k$, the undepreciated capital $1 - \delta$, and the marginal reduction in future adjustment costs $\phi(K_{t+2} - K_{t+1})$, all discounted by $\beta\lambda_{t+1}/\lambda_t$. The adjustment cost parameter $\phi$ prevents the capital stock from jumping instantaneously in response to shocks, generating the hump-shaped investment dynamics observed in the data \citep{Christiano2005}.

\subsection{Formal Sector (Tradable)}\label{sec:formal}

The formal sector is composed of a representative firm producing tradable goods. In this framework, we model AI not as a long-term capital stock, but as an imported intermediate input that the firm acquires in each period from the international market.

\subsubsection{Firm Set-up and Technology}

The formal sector is the locus of AI adoption in our model. Formal firms produce tradable goods ($Y_{f,t}$) using a nested CES--Cobb-Douglas technology that combines physical capital ($K_t$), formal labor $L_{f,t}$, and imported AI inputs $M_{ia,t}$:
\begin{equation}
Y_{f,t} = A_{f,t}\, K_t^{\alpha}\, Z_t^{1-\alpha}
\label{eq:Yf}
\end{equation}
where $A_{f,t}$ is total factor productivity in this sector, $\alpha \in (0,1)$ is the capital share, and $Z_t$ is a CES composite of imported AI capital and formal labor:
\begin{equation}
Z_t = \left[\theta_t\, M_{ia,t}^{\rho} + (1-\theta_t)\, L_{f,t}^{\rho}\right]^{1/\rho}, \qquad \rho = \frac{\sigma_{KL}-1}{\sigma_{KL}}
\label{eq:Z}
\end{equation}

The parameter $\rho$ is a transformation of the elasticity of substitution $\sigma_{KL}$ between AI capital and formal labor---\textit{the pivotal parameter of the model}.

\paragraph{The pivotal role of $\sigma_{KL}$.} The elasticity $\sigma_{KL} = 1/(1-\rho)$ measures how easily firms can substitute between AI capital and human labor when relative prices change. As \citet{AcemogluRestrepo2018} emphasize, this parameter governs whether automation \textit{displaces} or \textit{complements} workers:

When $\sigma_{KL} > 1$ ($\rho > 0$), AI and labor are \textit{gross substitutes}. A decline in the price of AI induces firms to replace workers with machines: the quantity of AI capital demanded rises more than proportionally, while demand for formal labor falls. This is the scenario that dominates public discourse about technological unemployment.

When $\sigma_{KL} < 1$ ($\rho < 0$), AI and labor are \textit{gross complements}. Cheaper AI increases the marginal product of labor, raising labor demand. Firms want \textit{more} of both inputs because they work better together. This is the scenario emphasized by researchers who view AI as augmenting, rather than replacing, human capabilities.

When $\sigma_{KL} = 1$ ($\rho \to 0$), the composite collapses to Cobb-Douglas, $Z_t = M_{ia,t}^{\theta} L_{f,t}^{1-\theta}$, and factor shares are constant regardless of relative prices.

The nesting structure in~\eqref{eq:Yf}--\eqref{eq:Z} is deliberate. The outer Cobb-Douglas layer between physical capital and the composite $Z_t$ imposes a unit elasticity of substitution between traditional capital and the AI--labor bundle, ensuring that the capital share $\alpha$ remains stable---consistent with the long-run evidence for most economies \citep{KarabarbounisNeiman2014}. All the action happens in the \textit{inner} CES layer, where the elasticity $\sigma_{KL}$ governs substitution between AI and workers. This separation allows the model to generate regime-switching behavior (through $\sigma_{KL}$) without altering the long-run capital share---a feature that single-layer CES models cannot deliver.

The empirical literature offers a range of estimates for capital-labor substitution elasticities. \citet{OberfieldRaval2021} estimate an aggregate elasticity of $0.5$--$0.7$ for U.S.\ manufacturing using plant-level data and find that it has declined slightly since 1970. However, their estimates pertain to traditional capital-labor substitution, not specifically to AI capital. For AI-specific substitution, no direct estimates exist for Latin America; the range $\sigma_{KL} \in [0.8, 1.6]$ that we explore is informed by indirect evidence from the occupational exposure literature \citep{Ciascchi2025}. Recent theoretical work by \citet{HemousOlsen2022} shows that even models with endogenous automation converge to a regime where the share of automating innovations increases over time, underscoring the importance of the substitution margin.

\subsubsection{Formal Sector Optimization Problem}

The representative firm maximizes instantaneous net profits, acting as a price taker for the formal wage $W_t$, the rental rate of capital $R_t^k$, and the exogenous international price of AI, $p_{ia,t}$:

\begin{equation}
\max_{\{K_{t}, L_{f,t}, M_{ia,t}\}} \Pi_{f,t} = Y_{f,t} - W_{t}L_{f,t} - R_{t}^{k}K_{t} - p_{ia,t}M_{ia,t}
\end{equation}
\\
subject to \eqref{eq:Yf} and \eqref{eq:Z}.

\subsubsection{First Order Conditions (FOCs)}

The capital demand is given by:
\begin{equation}
\frac{\partial \Pi_{f,t}}{\partial K_t} = \alpha A_{f,t} K_t^{\alpha-1} Z_t^{1-\alpha} - R_t^k = 0 \implies R_{t}^{k} = \alpha \frac{Y_{f,t}}{K_t}
\label{eq:Rk}
\end{equation}

This states that the rental rate of capital equals its marginal product. Since the outer layer is Cobb-Douglas, the capital share in formal output is constant at $\alpha$.

Using the chain rule, the marginal product of the composite $Z_t$ is:

\begin{equation}
\frac{\partial Y_{f,t}}{\partial Z_t} = (1-\alpha) A_{f,t} K_t^{\alpha} Z_t^{-\alpha} = (1-\alpha) \frac{Y_{f,t}}{Z_t}
\label{eq:composite}
\end{equation}

The marginal contributions of labor and AI to the composite $Z_t$ are:

\begin{equation*}
\frac{\partial Z_t}{\partial L_{f,t}} = Z_t^{1-\rho} (1-\theta_t) L_{f,t}^{\rho-1} \quad \text{and} \quad \frac{\partial Z_t}{\partial M_{ia,t}} = Z_t^{1-\rho} \theta_t M_{ia,t}^{\rho-1}
\label{eq:composite_L_AI}
\end{equation*}

Combining these results, the FOCs for the formal wage \eqref{eq:W_formal} and the price of AI \eqref{eq:p_ia} are:
\begin{align}
W_{t} &= (1-\alpha)\frac{Y_{f,t}}{Z_{t}}(1-\theta_{t})Z_{t}^{1-\rho}L_{f,t}^{\rho-1} \label{eq:W_formal} \\
p_{ia,t} &= (1-\alpha)\frac{Y_{f,t}}{Z_{t}}\theta_{t} Z_{t}^{1-\rho}M_{ia,t}^{\rho-1} \label{eq:p_ia}
\end{align}

Equations~\eqref{eq:W_formal} and~\eqref{eq:p_ia} jointly determine the demand for formal labor and AI capital given the wage, the AI price, and the level of output. Dividing~\eqref{eq:p_ia} by~\eqref{eq:W_formal} yields the relative demand condition:
\begin{equation}
\frac{p_{ia,t}}{W_t} = \frac{\theta_t}{1-\theta_t}\left(\frac{M_{ia,t}}{L_{f,t}}\right)^{\rho - 1}
\label{eq:relative_demand}
\end{equation}
This expression makes the regime-switching mechanism transparent. A decline in the relative price of AI raises the optimal AI-to-labor ratio. The key difference is how this adjustment affects labor demand. When \(\sigma_{KL}>1\), AI and labor are sufficiently substitutable for the increase in AI intensity to reduce formal labor demand. When \(\sigma_{KL}<1\), AI and labor are complements, so cheaper AI raises the marginal productivity of labor and can increase the demand for both inputs. Thus, the elasticity \(\sigma_{KL}\) governs whether higher AI intensity translates into labor displacement or labor augmentation.

\subsection{Informal Sector (Non-Tradable)}\label{sec:informal}

The informal sector represents the non-tradable goods market, characterized by a lack of access to advanced technological inputs. It serves as a labor buffer that absorbs displaced formal workers.

\subsubsection{Informal Firms Set--up and Technology}

Informal firms produce non-tradable goods using only labor with decreasing returns to scale:
\begin{equation}
Y_{i,t} = A_{i, t}\, L_{i,t}^{\gamma}, \quad \gamma \in (0,1)
\label{eq:Yi}
\end{equation}
where $A_{i,t}$ denotes informal-sector productivity and $\gamma$ governs returns to scale. The restriction $\gamma < 1$ reflects the well-documented fact that informal firms face diminishing marginal returns due to limited access to capital, technology, and organizational scale \citep{Ulyssea2018, LaPortaShleifer2014}.

The absence of capital in informal production is a simplification that we test in the robustness checks (Section~\ref{sec:results}). It captures the stylized fact that informal firms in Latin America are overwhelmingly small, labor-intensive, and capital-scarce. \citet{Ulyssea2018} documents using matched employer-employee data from Brazil that informal firms are substantially smaller and exhibit lower productivity than their formal counterparts. \citet{LaPortaShleifer2014} argue more broadly that informal firms are fundamentally different from formal ones---not merely formal firms evading taxes, but genuinely low-productivity operations run by less educated managers.

\subsubsection{Profit Maximization and First--Order Condition}

Informal firms maximize net profits $\Pi_{i,t}$ given by:
\begin{equation}
\max_{\{L_{i,t}\}} \Pi_{i,t} = p_{i,t}Y_{i,t} - W_{t}L_{i,t}
\end{equation}

We assume free labor mobility between sectors which implies that the household equalizes the return to working an additional hour in either sector. In the formal sector, this return is the wage $W_t$. In the informal sector, the return is the value of the marginal product of informal labor, $p_{i,t} \gamma A_{i, t} L_{i,t}^{\gamma-1}$. Setting these equal yields the \textit{labor mobility condition}:
\begin{equation}
W_t = p_{i,t}\,\gamma\, A_{i, t}\, L_{i,t}^{\gamma-1}
\label{eq:W_informal}
\end{equation}

This equation is the linchpin of the model's transmission mechanism. When the formal sector wage falls (say, because AI substitutes for labor), the left-hand side declines. To restore equality, informal employment $L_{i,t}$ must rise---driving down the marginal product of informal labor until the two sides are equalized again. This is precisely the \textit{buffering} role of the informal sector: it absorbs displaced workers when formal labor demand contracts, and releases workers when formal labor demand expands.

The decreasing returns parameter $\gamma$ governs the \textit{elasticity of this buffer}. When $\gamma$ is close to 1 (nearly constant returns), the informal sector can absorb large inflows of workers with only modest reductions in its marginal product; the buffer is elastic. When $\gamma$ is small (sharply decreasing returns), even small inflows push down the marginal product substantially; the buffer is rigid. We calibrate $\gamma = 0.65$, consistent with the labor share in informal GDP estimated from Latin American household surveys \citep{GaspariniTornarolli2009}.

\subsection{External Sector}\label{sec:external}

The economy borrows and lends in international capital markets at an interest rate that incorporates a country-specific risk premium. Following \citet{SchmittGroheUribe2003}, we specify:
\begin{equation}
r_t = r^* + \psi\left(e^{D_t - \bar{D}} - 1\right)
\label{eq:risk}
\end{equation}
where $r^*$ is the risk-free world interest rate, $\bar{D}$ is the steady-state level of external debt, and $\psi > 0$ governs the sensitivity of the risk premium to deviations of debt from its long-run level. When the country accumulates debt above $\bar{D}$, the exponential term rises above zero, increasing the borrowing cost. This specification serves two purposes: it ensures stationarity of the external debt process (a well-known requirement in small open economy models, as \citet{SchmittGroheUribe2003} demonstrate), and it captures the empirical regularity that emerging market spreads widen when external indebtedness rises \citep{UribeYue2006}.

\paragraph{Balance of payments.} The evolution of external debt is governed by:
\begin{equation}
D_{t+1}=(1+r_t)D_t+p_{ia,t}M_{ia,t}
-\left(Y_{f,t}-C_{f,t}-I_t-\frac{\phi}{2}(K_{t+1}-K_t)^2\right).
\label{eq:BOP}
\end{equation}

The right-hand side has three components. The first term, $(1+r_t)D_t$, captures debt service. The second term, $p_{ia,t}M_{ia,t}$, is the import bill associated with AI inputs. The final term captures the economy's net tradable surplus after domestic absorption, including formal consumption, physical investment, and capital adjustment costs. External debt increases when AI imports and domestic absorption exceed formal tradable output.

This equation reveals an important external adjustment channel. When the international price of AI falls, the import bill $p_{ia,t}M_{ia,t}$ may rise or fall depending on the price elasticity of AI demand. Under substitution $(\sigma_{KL}>1)$, the quantity of imported AI tends to rise strongly, so the volume effect may offset part of the price decline. Under complementarity $(\sigma_{KL}<1)$, AI demand is less elastic and the price effect is more likely to dominate. In general equilibrium, however, the external position also depends on the response of formal tradable output. In our baseline calibration, the expansion of formal output dominates the import-bill effect, so external debt declines under both regimes, with a stronger decline under complementarity. Thus, the sign of the external-debt response is a quantitative result rather than a mechanical implication of the substitution regime.

\subsection{Exogenous Shocks}\label{sec:shocks}

Four exogenous processes drive the model's dynamics:
\begin{align}
\ln A_{f,t} &= \rho_f\, \ln A_{f,t-1} + \epsilon_{f,t}, \quad \epsilon_{f,t} \sim N(0, \sigma_f^2) \label{eq:shock_A}\\[3pt]
\ln A_{i,t} &= \rho_i\, \ln A_{i,t-1} + \epsilon_{i,t}, \quad \epsilon_{i,t} \sim N(0, \sigma_i^2) \label{eq:shock_Ai}\\[3pt]
\ln p_{ia,t} &= (1\!-\!\rho_{ia})\ln\bar{p}_{ia} + \rho_{ia}\ln p_{ia,t-1} - \epsilon_{ia,t}, \quad \epsilon_{ia,t} \sim N(0, \sigma_{ia}^2) \label{eq:shock_p}\\[3pt]
\ln \theta_t &= (1-\rho_\theta)\ln \bar{\theta} + \rho_\theta \ln \theta_{t-1} + \epsilon_{\theta,t}, \quad \epsilon_{\theta,t} \sim N(0, \sigma_{\theta}^2)\label{eq:shock_theta_eq}
\end{align}

\section{Equilibrium}\label{sec:equilibrium}

A competitive equilibrium for this small open economy consists of a sequence of prices $\{W_t, R_t^k, p_{i,t}, r_t\}_{t=0}^{\infty}$ and allocations $\{C_t, C_{f,t}, C_{i, t}, L_t, L_{f, t}, L_{i, t}, I_t, K_t, D_{t+1}, M_{ia, t}, Y_{f, t}, Y_{i, t}, Z_t\}_{t=0}^{\infty}$ such that, given the initial conditions $\{K_0, D_0\}$ and exogenous processes $\{A_{f, t}, A_{i, t}, \theta_t, p_{ia, t}\}_{t=0}^{\infty}$, households and firms maximize their objective functions and all markets clear. The dynamical system is characterized by the following blocks of equations:

\subsection{Households and Intertemporal Choice}

Substituting the consumption first-order condition into the labor-supply condition yields the aggregate labor-market equilibrium condition:

\begin{equation}
 \chi L_t^\nu P_t = W_t
 \label{eq:equilibium_l}
\end{equation}

This condition equates the marginal disutility of total labor supply, expressed in units of the consumption bundle, to the nominal value of the wage measured in units of the tradable numéraire. Because formal and informal hours enter symmetrically in the household's disutility of labor, aggregate labor supply depends on total hours worked, while the allocation of those hours across sectors is determined by firm-side labor demand and the wage-equalization condition.

Plug in \eqref{eq:FOC_C} in \eqref{eq:FOC_D} we obtain the Euler equation for the debt (Equation \eqref{eq:debt_euler}). This equation is the standard intertemporal Euler equation governing the household's consumption-saving trajectory. It dictates that, along the optimal path, the marginal utility of forgone consumption today (the left-hand side) must exactly equal the expected, discounted marginal utility of the augmented consumption tomorrow (the right-hand side). In the context of our small open economy framework, this condition illustrates how the household smooths aggregate consumption by trading one-period international bonds.

\begin{equation}
    \frac{\left(C_t - \chi \frac{L_t^{1+\nu}}{1+\nu}\right)^{-\sigma}}{P_t} = \beta (1+r_t) \mathbb{E}_t \left[\frac{\left(C_{t+1} - \chi \frac{L_{t+1}^{1+\nu}}{1+\nu}\right)^{-\sigma}}{P_{t+1}} \right]
    \label{eq:debt_euler}
\end{equation}

Plug in \eqref{eq:FOC_C} in \eqref{eq:FOC_K} we obtain the intertemporal optimality condition for physical capital accumulation (Equation \eqref{eq:euler_K}). The left-hand side captures the marginal cost of investing an additional unit of capital today—evaluated in terms of marginal utility—which includes both the direct cost of forgone consumption and the marginal capital adjustment cost governed by $\phi$. The right-hand side represents the expected, discounted marginal benefit of entering tomorrow with that extra unit of capital. This return comprises the future rental rate $R_{t+1}^k$, the undepreciated capital $1-\delta$, and the marginal savings on tomorrow's adjustment costs. 

\begin{equation}
    \frac{\left(C_t - \chi \frac{L_t^{1+\nu}}{1+\nu}\right)^{-\sigma}}{P_t} [1 + \phi(K_{t+1} - K_t)] = \beta \mathbb{E}_t \left[ \frac{\left(C_{t+1} - \chi \frac{L_{t+1}^{1+\nu}}{1+\nu}\right)^{-\sigma}}{P_{t+1}} (R_{t+1}^k + 1 - \delta + \phi(K_{t+2} - K_{t+1})) \right]
    \label{eq:euler_K}
\end{equation}

So, the presence of these quadratic adjustment costs ensures that the physical capital stock evolves gradually rather than jumping instantaneously. In the context of our model, this friction is vital: because physical capital cannot adjust immediately to an AI price shock, the labor market—and thereby the informal sector—is forced to act as the primary buffer for short-term macroeconomic adjustments.

\subsection{Intratemporal Consumption Allocation}

By minimizing the cost of the aggregate consumption bundle subject to the relative price of informal goods, we obtain the household's optimal demand ratio for formal versus informal consumption. This relative price is determined by the household's intratemporal allocation as defined in \eqref{eq:relprice}.

\subsection{Firms and Factor Demands}

In equilibrium, factor prices must equal their respective marginal products. For the formal sector, the optimal demands for physical capital, formal labor, and imported AI are strictly determined by equations \eqref{eq:Rk}, \eqref{eq:W_formal}, and \eqref{eq:p_ia}. Simultaneously, the informal sector's labor demand is governed by the wage equalization condition \eqref{eq:W_informal}, which ensures that the formal wage equals the value of the marginal product of informal labor.

\subsection{Market Clearing and Aggregate Resource Constraint}

General equilibrium requires that all domestic markets clear. The non-tradable informal goods market clears when domestic production is fully consumed internally:
\begin{equation}
    Y_{i,t} = C_{i,t}
    \label{eq:clearing_i}
\end{equation}
This closes the informal sector block. Combined with the relative price equation~\eqref{eq:relprice} and the labor mobility condition~\eqref{eq:W_informal}, it determines the three informal sector variables---$Y_{i,t}$, $L_{i,t}$, and $p_{i,t}$---as functions of the wage and tradable consumption.
Similarly, the aggregate labor market clears when the total labor supplied by households is fully absorbed by the competing demands of the two sectors:
\begin{equation}
    L_t = L_{f,t} + L_{i,t}
    \label{eq:market_clearing_L}
\end{equation}

To derive the economy-wide resource constraint, we consolidate the household's budget constraint with the profit functions of both formal and informal firms. By imposing the non-tradable market clearing condition \eqref{eq:clearing_i}, the domestic non-tradable terms cancel out, yielding the aggregate resource constraint for the tradable sector:
\begin{equation}
    C_{f, t} + I_t + \frac{D_{t+1}}{1+r_t} + \frac{\phi}{2}(K_{t+1}-K_t)^2 = D_t + Y_{f,t} - p_{ia,t} M_{ia,t}
    \label{eq:resource_constraint}
\end{equation}

The shortfall between domestic tradable production and total domestic absorption dictates the economy's external financing needs. We rewrite \eqref{eq:resource_constraint} as follows:
\begin{equation}
    D_{t+1} = (1+r_t) D_t + p_{ia,t} M_{ia,t} - \left( Y_{f,t} - C_{f,t} - I_t - \frac{\phi}{2}(K_{t+1}-K_t)^2 \right)
    \label{eq:BOP_equilibrium}
\end{equation}

This external debt trajectory is coupled with an External Debt-Elastic Interest Rate (EDEIR) rule (Equation \eqref{eq:risk}). Following \citet{SchmittGroheUribe2003}, this specification incorporates a country risk premium sensitive to the aggregate debt level, thereby ensuring the stationarity of the small open economy framework and closing the model around a well-defined steady state.

Finally, to rule out explosive borrowing paths and ensure that the household's intertemporal budget constraint binds in the long run, the optimal plan must satisfy the No-Ponzi game condition for external debt:
\begin{equation}
    \lim_{t \to \infty} \mathbb{E}_0 \left[ \frac{D_{t+1}}{\prod_{s=0}^{t} (1+r_s)} \right] = 0
    \label{eq:no_ponzi}
\end{equation}
Similarly, the optimal accumulation of physical capital is bounded by the standard transversality condition, which ensures that the discounted shadow value of the capital stock approaches zero in the limit:
\begin{equation}
    \lim_{t\to\infty}\beta^t E_0[\lambda_t K_{t+1}]=0, \qquad
    \lambda_t= \frac{\left(C_t-\chi\frac{L_t^{1+\nu}}{1+\nu}\right)^{-\sigma}}{P_t}
    \label{eq:tvc_K}
\end{equation}

\subsection{Summary of the Model's Mechanism}

The model is solved by log-linearizing the equilibrium system around the deterministic steady state using first-order perturbation methods in Dynare. The core mechanism can be summarized as follows. A decline in the international price of AI capital triggers a formal-sector adjustment whose direction depends on $\sigma_{KL}$. Under substitution ($\sigma_{KL}>1$), cheaper AI puts partial-equilibrium downward pressure on formal labor demand, inducing labor reallocation toward the informal sector. However, the aggregate wage response is not determined by this substitution channel alone. In general equilibrium, lower AI costs also affect formal output, capital accumulation, relative prices, and non-tradable demand. As a result, the aggregate real wage may increase even when formal employment contracts. Under complementarity ($\sigma_{KL}<1$), cheaper AI raises the marginal productivity of formal labor, expands formal employment, and amplifies output, wages, investment, and capital accumulation. The model therefore shows that $\sigma_{KL}$ is pivotal for the direction of labor reallocation, while the aggregate responses of wages, informality, external debt, and relative prices depend on broader general-equilibrium feedbacks operating through demand-side substitutability, relative prices, capital accumulation, and the external sector.

%% ============================================================
\section{Related Empirical Evidence}\label{sec:empirical}
%% ============================================================

The model posits that everything hinges on whether AI complements or substitutes for labor. This section reviews recent empirical work to assess whether data patterns align with our mechanisms.

\subsection{Methodology in the Literature}

Researchers measure AI exposure at the occupation level using expert assessments or language model--based scoring, then link these to individual-level household surveys \citep{AcemogluRestrepo2018}. Four indices are now commonly used in this literature: the Artificial Intelligence Occupational Exposure index (AIOE), the Complementarity-Adjusted Artificial Intelligence Occupational Exposure index (C-AIOE), the Generative AI Exposure index developed by Gmyrek, Berg, and Bescond (GBB), and the AI-Generated Occupational Exposure index (GENOE) \citep{Ciascchi2025}.

\subsection{Key Findings}

Between 26 and 38 percent of Latin American jobs could experience influence from Generative AI, but only 2--5 percent face serious automation risk \citep{ILO2024}. Informality dramatically reduces measured exposure: formal workers face much higher potential exposure to both automation and augmentation. Exposure concentrates among women, younger workers, educated workers, and formal employees \citep{Ciascchi2025, Azuara2024}. About 17 million positions that could see productivity gains belong to workers who lack adequate digital access \citep{ILO2024}.

\subsection{Implications for Our Model}

The evidence lines up with our framework: AI exposure is real but uneven, informality reduces exposure, exposure concentrates among formal workers, digital infrastructure matters, and the complementarity-vs-substitution distinction changes the distributional picture. The simulation evidence from \citet{Ciascchi2025}---that complementarity-adjusted measures show different patterns---directly supports our claim that $\sigma_{KL}$ is the key parameter.

\subsection{AI Adoption and Traditional Productivity Shocks}\label{sec:other_shocks}

While the AI price shock isolates the pure cost effect of imported technology, the model also accommodates shifts in AI adoption intensity ($\theta_t$) and traditional total factor productivity in both sectors ($A_{f,t}$, $A_{i,t}$). Analyzing these shocks clarifies why the elasticity of substitution plays such a unique role in the context of price fluctuations.

An exogenous increase in the AI share parameter ($\theta_t$)---reflecting a management-driven or organizational bias toward automation---operates differently than a price decline. Rather than moving the firm along a given isoquant in response to cheaper inputs, a positive shock to $\theta_t$ represents a structural shift in the production technology heavily biased toward AI capital. Under the substitution regime, this triggers a sharp, immediate contraction of formal employment. Crucially, because the economy must still import this AI capital at the prevailing international price, this "bias-driven" adoption generates a more severe deterioration of the trade balance and a sharper accumulation of external debt than a price-driven adoption. This dynamic captures the risk of "excessive automation" \citep{AcemogluRestrepo2018}: premature adoption of AI in emerging markets can hollow out the formal sector and drain foreign reserves without delivering the offsetting benefits of cheaper production costs.

Conversely, standard neutral productivity shocks ($A_{f,t}$) behave conventionally and do not hinge on the $\sigma_{KL}$ threshold. A positive formal TFP shock raises the marginal product of all factors simultaneously. This increases formal labor demand and wages, drawing workers out of the informal sector and reducing the informality rate. Furthermore, the expansion of tradable output generates a Balassa-Samuelson-type effect: the relative abundance of formal goods drives up the relative price of non-tradables ($p_{i,t}$), generating a real exchange rate appreciation that further boosts the value of the marginal product in the informal sector, restoring equilibrium at a higher aggregate wage level.

Finally, a positive shock to informal productivity ($A_{i,t}$)---such as the diffusion of digital payment platforms among unregistered businesses---acts effectively as a negative labor supply shock to the formal economy. By raising the marginal product of informal labor, it increases the workers' outside option. To satisfy the mobility condition \eqref{eq:W_informal}, formal firms are forced to pay higher wages, which squeezes formal profits and mildly contracts formal employment, illustrating the two-way transmission channel inherent to dual labor markets. 

This contrast is fundamental: it confirms that the regime-switching behavior documented in Section~\ref{sec:mechanism} is not a generic feature of macroeconomic volatility, but a unique property of AI-specific cost shocks interacting with the substitution elasticity. Detailed impulse response functions for these secondary shocks are provided in Appendix C.

%% ============================================================
\section{Calibration and Solution}\label{sec:calibration}
%% ============================================================

\subsection{Steady State}\label{sec:steadystate}

The steady state is solved semi-analytically. A pivotal property of the model is that the formal wage $W$ is independent of the scale of employment, determined as a fixed-point $W = f(W)$ as shown in Appendix B.1. From the Euler equations:
\begin{equation}
R^k = \frac{1}{\beta} - 1 + \delta, \qquad r = \frac{1}{\beta} - 1
\label{eq:ss_rates}
\end{equation}

A key analytical result of our framework is that the formal sector real wage, $W$, is independent of the scale of employment in the steady state. As derived in Appendix B, the marginal productivity conditions of the formal firm can be collapsed into a single fixed-point equation for the wage:
\begin{equation}
W = (1-\alpha)(1-\theta_t)\left(\frac{\alpha}{R^k}\right)^{\!\alpha/(1-\alpha)} \!\! g(x)^{1-\rho}
\label{eq:fixed_point_W}
\end{equation}
where $x = M_{ia}/L_f = [\theta_t W / ((1-\theta_t)p_{ia})]^{1/(1-\rho)}$ represents the AI-to-labor ratio and $g(x) = [\theta_t x^{\rho} + (1-\theta_t)]^{1/\rho}$ is the intensive-form CES composite. We solve this system using a numerical shooting algorithm that converges rapidly.

Once $W$ is identified, the rest of the model follows a chain of determination. First, the labor mobility condition \eqref{eq:W_informal} ensures the informal sector matches this wage. Second, the aggregate resource constraint \eqref{eq:resource_constraint} in steady state dictates the economy's trade balance requirement:
\begin{equation}
Y_f - C_f - \delta K = r \bar{D} + p_{ia} M_{ia}
\label{eq:ss_trade}
\end{equation}
This condition underscores that in the long run, the consumption of tradables and the maintenance of the capital stock are constrained by the service of external debt and the ongoing bill for imported AI capital. Finally, the labor disutility parameter $\chi$ is residually determined to ensure the aggregate labor supply condition \eqref{eq:equilibium_l} holds at the targeted employment levels.

\subsection{Calibration}

We calibrate the model at a quarterly frequency to represent a typical Latin American economy. First, we set the coefficient of relative risk aversion to two ($\sigma = 2$), implying an intertemporal elasticity of substitution of $0.5$. This value is consistent with a broad range of estimates for emerging economies \citep{AguiarGopinath2007}.

Second, we calibrate the baseline value of the intratemporal elasticity of substitution between tradable and non-tradable goods at $\eta=0.95$, a value close to the Cobb--Douglas case $(\eta=1)$. This calibration provides a baseline in which households can substitute between the two types of goods, but only to a limited extent. As a result, neither strong complementarity nor strong substitutability dominates the transmission mechanism, allowing us to isolate more clearly the effects of AI adoption operating through production decisions, labor reallocation, and external adjustment. To assess the robustness of our results, we also consider two alternative values. First, we set $\eta=0.44$, following \citet{OstryReinhart1992}, which implies a high degree of complementarity between tradable and non-tradable goods. In this case, productivity gains in the formal sector generate demand spillovers toward the informal sector, reinforcing positive linkages across sectors. Second, we consider $\eta=1.50$, which implies greater substitutability between tradable and non-tradable goods. Under this calibration, households can adjust their expenditure composition more easily in response to relative price movements, making expenditure-switching effects more important than demand complementarities. Together, these alternative calibrations span a plausible range of values for developing economies and allow us to evaluate the sensitivity of the model's predictions to different assumptions regarding consumption substitutability. Appendix \ref{app:secondary_consumption} reports the simulations under the alternative calibrations.

Third, we set the depreciation rate to $\delta = 0.025$ at quarterly frequency, which corresponds to an annual depreciation rate of approximately 10 percent. Fourth, we set $\nu = 1.5$, implying a Frisch elasticity of labor supply of $1/\nu \approx 0.67$, which lies within the range reported by \citet{SmetsWouters2007}. 

Fifth, we calibrate the informal sector productivity $A_i$ to match an informality rate ($L_i/L$) of 50\%, consistent with Latin American averages reported by \citet{GaspariniTornarolli2009}. Sixth, we set the AI adoption parameter to $\bar\theta = 0.1$ reflecting that AI investment remains a small share of total factor payments in Latin America. Following \citet{SchmittGroheUribe2018}, we calibrate $\psi = 5.49$, which corresponds to the median debt elasticity of the interest rate estimated for a sample of 13 Latin American economies. We set the discount factor to $\beta = 0.98$, implying an annual real interest rate of approximately 8 percent, a value commonly used in small open economy models for emerging markets. The capital share in formal-sector production is calibrated to $\alpha = 0.40$, consistent with national accounts evidence indicating that labor shares in Latin American economies are generally lower than those observed in advanced economies. We set the weight of tradable goods in the consumption basket to $\omega = 0.4$, reflecting the relatively large role of non-tradable goods and services in household expenditure. The returns-to-scale parameter in the informal sector is calibrated to $\gamma = 0.65$, implying diminishing marginal productivity and capturing the lower efficiency and smaller scale typically associated with informal production activities. Finally, we set the investment adjustment cost parameter to $\phi = 4.96$, following \citet{GarciaCicco2010}.

\subsubsection{Plausible Values for $\sigma_{KL}$}

Direct estimates of $\sigma_{KL}$ between AI capital and labor are not available for Latin America. \citet{Ciascchi2025} construct a Complementarity-Adjusted AIOE (C-AIOE) index; occupations with flatter gradients across income distributions suggest $\sigma_{KL} < 1$, while steeper gradients suggest $\sigma_{KL} > 1$. The range $\sigma_{KL} \in [0.8, 1.6]$ encompasses the 10th--90th percentiles of implied values. We calibrate $\sigma_{KL} = 1.5$ (substitution) and $\sigma_{KL} = 0.8$ (complementarity). Sensitivity analysis in Section~\ref{sec:sensitivity} confirms robustness within $[0.5, 2.0]$.

\subsubsection{Toward an Observable AI Capital Price Index}\label{sec:priceindex}

The model treats $p_{ia,t}$ as an exogenous AR(1) process with $\rho_{ia} = 0.9$ and $\sigma_{ia} = 0.01$, normalized to unity in steady state. The quantitative content would be strengthened by grounding these in observable data. Three categories of publicly available data could form the basis of such an index:

\textit{GPU prices.} Nvidia's quarterly financial reports (10-K and 10-Q filings, available at \url{https://investor.nvidia.com}) provide revenue and unit shipment data for data center GPUs. Average selling prices can be inferred, and third-party trackers provide supplementary estimates. A quarterly series from 2018Q1 to 2025Q4 is constructible.

\textit{Cloud computing costs.} Amazon Web Services, Microsoft Azure, and Google Cloud publish list prices for compute instances. Research groups have assembled historical price series for standardized GPU-hour workloads.

\textit{Capital goods deflators.} CEPALSTAT and national accounts provide import price deflators for capital goods categories that include computing equipment, available at quarterly frequency for most Latin American economies.\footnote{See \url{https://statistics.cepal.org}.}

The secular decline in AI-capital prices is well documented in the semiconductor literature. For instance, the price per transistor has followed a roughly exponential decline since the 1970s, while GPU price-performance ratios have improved rapidly in recent years. However, a formal quarterly price series specific to AI-capital imports in Latin America has not yet been constructed. We therefore calibrate the AI-price process rather than estimate it directly. The persistence parameter $\rho_{ia}=0.9$ captures the gradual nature of technology diffusion, while $\sigma_{ia}=0.01$ is a conservative calibration intended to represent modest quarterly fluctuations around the trend. Transitory disruptions, such as the 2021--2022 semiconductor shortage, can be approximated within this AR(1) structure.

Constructing even a coarse quarterly AI-capital import price index based on GPU average selling prices, cloud GPU-hour list prices, and capital-goods deflators from CEPALSTAT would allow future work to re-estimate $\rho_{ia}$ and $\sigma_{ia}$ using observed price variation. This would move the AI-price block from calibration toward estimation. We therefore view the construction of such an index as a central empirical agenda for companion work and as a key input for the structural identification of the relevant substitution elasticity $\sigma_{KL}$ in Latin American data. Table~\ref{tab:params} summarizes the calibrated parameters.

\begin{table}[H]
\centering
\caption{Calibrated Parameters}\label{tab:params}
\small
\begin{tabular}{@{}lll@{}}
\toprule
\textbf{Parameter} & \textbf{Value} & \textbf{Description / Source} \\
\midrule
\multicolumn{3}{@{}l}{\textit{Households}} \\
$\beta$ & 0.98 & Discount factor (8\% annual real rate) \\
$\sigma$ & 2.0 & Risk aversion (standard) \\
$\nu$ & 1.5 & Inverse Frisch elasticity \citep{SmetsWouters2007} \\
$\chi$ & 1.0 & Labor disutility (calibrated in SS) \\
$\omega$ & 0.4 & Tradable goods weight \\
$\eta$ & 0.95 & Intratemporal elasticity (Baseline) \\
\addlinespace
\multicolumn{3}{@{}l}{\textit{Formal Sector}} \\
$\alpha$ & 0.40 & Capital share \\
$\bar\theta$ & 0.1 & AI share in composite \\
$\sigma_{KL}$ & 1.5 / 0.8 & Substitution elasticity (Section 7.2.1) \\
$\delta$ & 0.025 & Depreciation (10\% annual) \\
$\phi$ & 4.96 & Adjustment cost \citep{GarciaCicco2010} \\
$A_f$ & 1 & Normalized at steady-state\\
\addlinespace
\multicolumn{3}{@{}l}{\textit{Informal Sector}} \\
$\gamma$ & 0.65 & Returns to scale \\
$A_i$ & 1 & Normalized at steady-state\\
\addlinespace
\multicolumn{3}{@{}l}{\textit{External Sector}} \\
$r^*$ & 0.0204 & International rate (from $\beta$) \\
$\psi$ & 5.49 & Risk premium \citep{SchmittGroheUribe2018} \\
$\bar{D}$ & 0.1 & SS debt \\
$\bar{p}_{ia}$ & 1.0 & Normalized AI price \\
\addlinespace
\multicolumn{3}{@{}l}{\textit{Shocks}} \\
$\rho_f$, $\rho_{ia}$ & 0.9 & Persistence \citep{SmetsWouters2007} \\
$\sigma_f$, $\sigma_{ia}$ & 0.01 & Std.\ deviations \\
\bottomrule
\end{tabular}
\end{table}

% \subsection{Solution Method}

% We log-linearize around the deterministic steady state and solve using first-order perturbation in Dynare. The Blanchard-Kahn conditions are satisfied. 
% The complete code is in Appendix~\ref{app:dynare}.

%% ============================================================
\section{Results}\label{sec:results}
%% ============================================================

\subsection{Impulse Responses to an AI Price Shock}

In this section, we present the quantitative analysis of the model. We first examine the dynamics following a persistent 1\% decline in the international price of AI capital under our baseline balanced-demand configuration.\footnote{This baseline configuration fixes $\eta = 0.95$. The alternative demand-side regimes highlighting high consumption complementarity ($\eta = 0.44$) and high consumption substitutability ($\eta = 1.50$) are analyzed in Appendix \ref{app:secondary_consumption}.} We then exploit our mechanism identification strategy to demonstrate how non-linear general equilibrium interactions between supply-side technology ($\sigma_{KL}$) and demand-side preferences ($\eta$) govern labor reallocation and wage dynamics in developing dual labor markets.

\subsubsection{Substitution Regime ($\sigma_{KL} = 1.5$)}

In the substitution regime, artificial intelligence and formal human labor act as gross substitutes within the nested production function of the tradable sector. Following a persistent negative shock to the international price of artificial intelligence, the primary transmission mechanism is driven by the relative factor demand condition. A cheaper international price of automation technology alters the relative input price ratio, prompting formal firms to optimize their intensive-form cost structure by substituting human labor with imported technology. As a direct microeconomic consequence of this task-displacement effect, formal employment contracts immediately on impact. 

Because the labor market operates under competitive, frictionless neoclassical conditions, the formal sector adjustment forces an instantaneous reallocation of the domestic workforce. Displaced formal workers migrate toward the informal sector, which acts as an endogenous macroeconomic buffer by absorbing the labor surplus. This labor supply push expands total informal hours worked, leading to a subsequent expansion in the production of non-tradable goods. This transition illustrates the safety-net role played by informal economic activities in emerging markets when facing labor-saving technology shocks.

Despite the immediate contraction in formal hours, the economy-wide aggregate real wage reacts positively. This outcome is governed by a tension between the partial-equilibrium factor substitution inside the inner production composite and the general-equilibrium scale effect. The price reduction of the imported intermediate input lowers the formal firm's marginal cost, triggering an expansion in total tradable output. Simultaneously, the consumption block incorporates an intratemporal substitution layer where formal and informal goods are imperfect substitutes. The expansion of formal sector income increases aggregate household wealth, shifting consumer demand toward non-tradables. Due to this demand bias, the relative price of informal goods experiences an appreciation that elevates the marginal revenue product of informal labor, successfully insulating the real wage from the downward pressure of the labor supply inflow.

The external sector and investment choices complete the adjustment loop. Although the quantity of imported artificial intelligence expands, the drop in its international unit price dominates the volume effect, causing the overall AI import bill to contract. The trade balance improves, allowing the representative household to smooth consumption by systematically drawing down its stock of external debt. Under the debt-elastic interest rate premium rule, this persistent deleveraging process reduces the country risk premium and lowers the domestic borrowing cost. The relaxed interest rate environment reduces the user cost of capital, stimulating physical investment and driving a gradual, hump-shaped accumulation of traditional physical capital. Because physical capital and the automated composite are bound by a unit-elastic Cobb-Douglas layer, this capital-deepening process increases the marginal productivity of formal workers over the medium-term transition, successfully mitigating the initial technological displacement effect.

\subsubsection{Complementarity Regime ($\sigma_{KL} = 0.8$)}

\begin{figure}[H]
\centering
\makebox[\textwidth][c]{\includegraphics[width=1\textwidth]{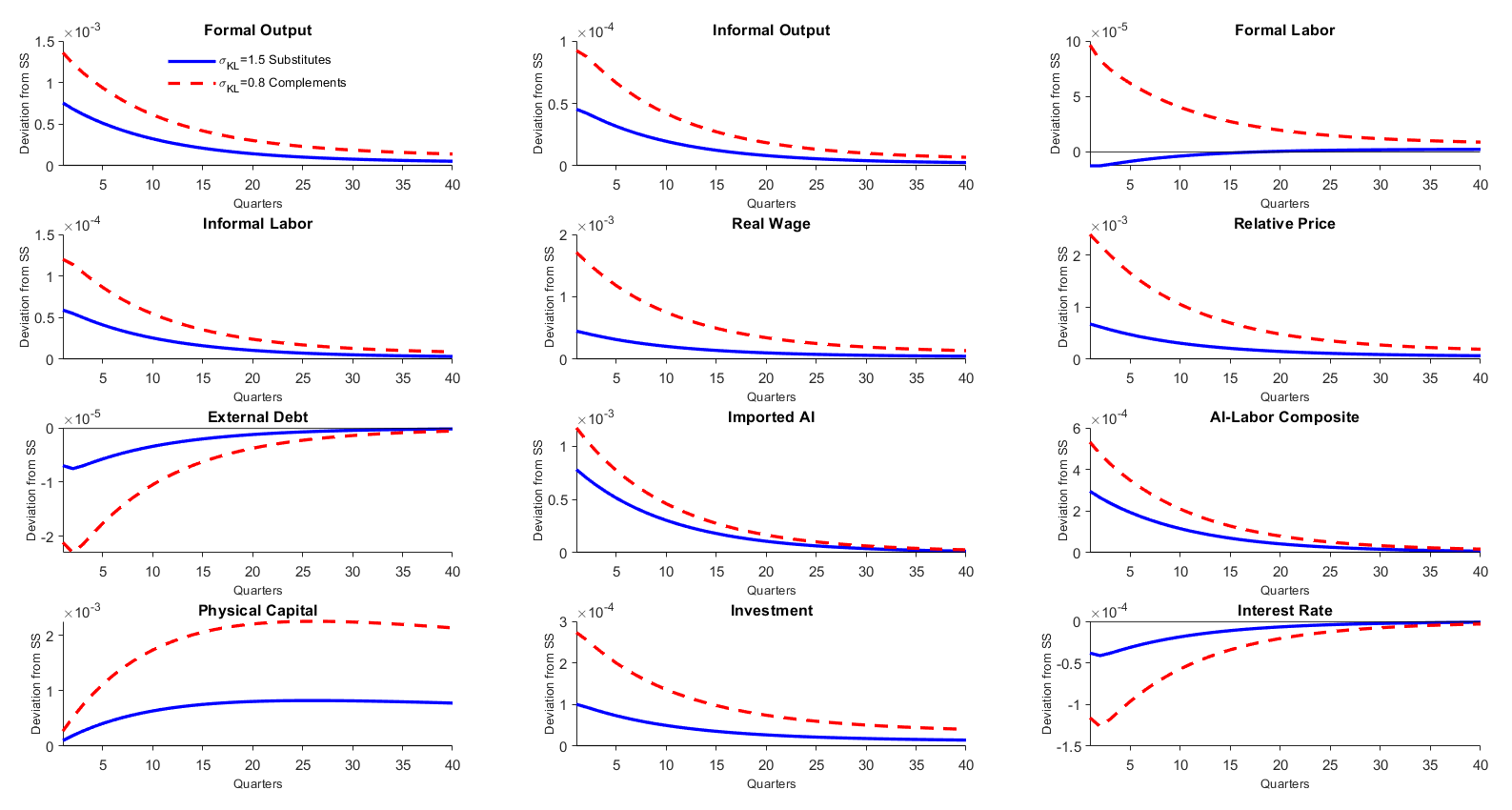}}
\caption{Impulse responses to a 1\% decline in the international price of AI. Blue solid lines represent the substitution regime ($\sigma_{KL} = 1.5$) and red dashed lines represent the complementarity regime ($\sigma_{KL} = 0.8$).}
\label{fig:irfs}
\end{figure}

In the complementarity regime, where $\sigma_{KL} = 0.8$, the decline in the international price of artificial intelligence does not displace workers, but instead directly increases the marginal productivity of human tasks within the formal sector. Because AI inputs and human labor act as gross complements inside the nested production composite, the reduction in imported technology costs functions as a direct efficiency gain for formal firms. This synergistic interaction triggers an expansion in formal output, which drives an immediate increase in formal employment upon impact. This positive labor market response stands in sharp contrast to the substitution scenario, demonstrating that the scale effect under technological complementarity overpowers the factor-saving channel. This expansionary path is amplified by a powerful capital-deepening loop. Investment and the traditional physical capital stock show positive responses with amplitudes that double those observed in the substitution scenario. Since physical capital and the automated composite are combined via a unit-elastic Cobb-Douglas layer, the rapid accumulation of machinery further elevates the marginal product of formal workers, reinforcing the initial technology shock over the transition. This production boom causes the economy-wide real wage to rise sharply, reaching magnitudes larger than those of the substitution case. Under the Greenwood, Hercowitz, and Huffman preference specification, this real wage surge motivates households to expand the total supply of labor hours in the economy. In this context, the growth of the aggregate labor force expands the total pool of hours enough to relax the traditional inter-sectoral trade-off. Consequently, the formal sector can increase its workforce without emptying the informal sector. Informal employment also expands to satisfy the rising demand for non-tradable goods, which is pulled by the increase in the total income of the economy and the subsequent appreciation of the relative price of informal services. Finally, the external adjustment stabilizes the macroeconomic transition. The formal output expansion allows the economy to improve its trade balance, leading to a persistent decline in external debt. Through the debt-elastic risk premium rule, this deleveraging process reduces the country risk premium and drives a substantial fall in the domestic interest rate, lowering borrowing costs and sustaining the physical capital expansion over the medium term (Figure \ref{fig:irfs}). 

\subsection{Mechanism Identification: The Interaction Between Technology and Demand Elasticities}\label{sec:mechanism}

\begin{figure}[H]
\centering
\includegraphics[width=0.85\textwidth]{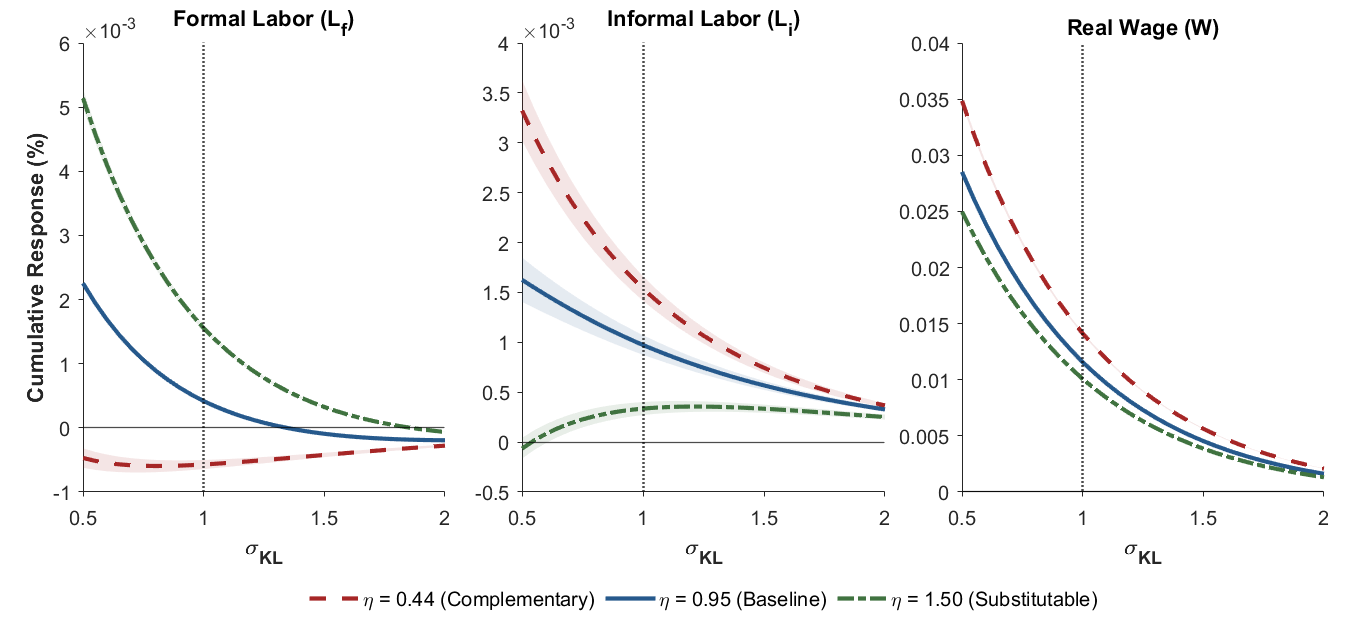}
\caption{Cumulative impulse responses over 20 quarters as functions of $\sigma_{KL}$ across alternative consumer preferences ($\eta$). Vertical dotted line marks the partial equilibrium Cobb-Douglas boundary ($\sigma_{KL} = 1$). Shaded bands represent the 10th--90th percentile sensitivity ranges generated by varying the informal returns-to-scale parameter.}
\label{fig:sigma}
\end{figure}

To further unpack the general-equilibrium transmission of an international AI price shock, we conduct a joint parametric sweep. Figure~\ref{fig:sigma} plots the cumulative 20-quarter responses of formal employment, informal employment, and real wages as continuous functions of the technology elasticity, $\sigma_{KL}\in[0.5,2.0]$, under three goods-market regimes: consumption complementarity ($\eta=0.44$), the baseline near Cobb--Douglas case ($\eta=0.95$), and consumption substitutability ($\eta=1.50$).

A key result from this exercise is that the general-equilibrium threshold does not coincide with the partial-equilibrium Cobb--Douglas benchmark. In partial equilibrium, one would expect the sign of the formal-employment response to switch at $\sigma_{KL}=1$. In general equilibrium, however, the zero-crossing occurs at a higher value. Under the baseline calibration with $\eta=0.95$, formal employment changes sign at approximately $\sigma_{KL}=1.35$. Under high consumption substitutability, $\eta=1.50$, the threshold shifts further to approximately $\sigma_{KL}=1.88$. This displacement of the threshold reflects the role of capital deepening and demand reallocation, which allow the scale effect to dominate even within part of the substitution region.

Under the baseline and high-substitutability goods-market regimes, a decline in $p_{ia,t}$ operates as a formalization force over a wide range of values of $\sigma_{KL}$. The fall in imported technology costs stimulates AI adoption and raises the productivity of the formal sector. Through general-equilibrium feedbacks, this also encourages physical capital accumulation, increasing the marginal productivity of the AI--labor composite $Z_t$. As a result, the scale effect can dominate the direct displacement effect even when AI and formal labor are moderate technological substitutes.

When technological substitutability becomes sufficiently high, the displacement effect eventually dominates under the baseline calibration. In that region, formal labor contracts and informal employment increases, consistent with the role of the informal sector as an employment buffer. Under higher consumption substitutability, however, expenditure reallocation toward formal tradable goods weakens the expansion of informal demand, so the informal sector absorbs less labor and the formalization effect remains stronger over a broader range of $\sigma_{KL}$.

The mechanism changes sharply when formal and informal goods are strong complements in the consumption basket ($\eta=0.44$). In this case, the expansion of formal output generated by cheaper AI also raises demand for informal non-tradable goods. Since the informal sector uses a labor-only technology with decreasing returns, satisfying this additional demand requires a stronger increase in informal labor and a rise in the relative price of informal goods, $p_{i,t}$. Through the wage-equalization condition~\eqref{eq:W_informal}, this relative-price adjustment puts upward pressure on the economy-wide wage. The resulting wage pressure can offset the technological incentives to expand formal employment, so the cumulative response of formal labor is weaker, and may turn negative, even for values of $\sigma_{KL}$ that would otherwise favor formal-sector expansion.

\paragraph{Decomposing the result: CES properties vs.\ model contributions.}

It is important to distinguish what is a mechanical property of the CES production function from what emerges from the model's structure. The sign of the regime switch---factor demands reversing at $\sigma_{KL} = 1$---is a well-known property of any CES technology, established since \citet{ACMS1961}. Any model with a CES composite will produce opposite responses depending on whether the elasticity exceeds or falls below unity. This property alone is not novel.

What the model contributes beyond this operates through three specific channels. First, the \textit{informality transmission channel}: the dual labor market translates the formal-sector demand shift into a quantifiable change in the informality rate via the mobility condition~\eqref{eq:W_informal}, with the magnitude governed by $\gamma$. This channel is absent from single-sector models. Second, the \textit{external debt channel}: because AI capital is imported, the price shock enters the balance of payments~\eqref{eq:BOP}, generating dynamics between technology adoption, trade balance, and country risk. Third, the \textit{relative price channel}: changes in informal output alter the relative price of non-tradables, creating a real exchange rate adjustment that feeds back into consumption and welfare. These three channels interact simultaneously; their combined effect cannot be derived from the CES alone.

%% ============================================================
\subsection{Sensitivity Analysis and Global Robustness}\label{sec:sensitivity}
%% ============================================================

To validate the structural stability of our baseline general equilibrium responses against parameter uncertainty, Table~\ref{tab:sensitivity} reports the cumulative boundaries extracted directly from our three-dimensional parametric sweep. We report the exact ranges over a 20-quarter horizon while randomizing the informal sector's returns-to-scale parameter ($\gamma \in [0.55, \; 0.75]$) under our baseline balanced-demand configuration ($\eta = 0.95$). To maintain visual and numerical clarity given the smooth perturbation scale of the small open economy framework.

\begin{table}[H]
\centering
\caption{Sensitivity Analysis: Cumulative Responses}\label{tab:sensitivity}
\small
\begin{tabular}{@{}lcc@{}}
\toprule
\textbf{Outcome} & \textbf{Substitution ($\sigma_{KL}=1.5$)} & \textbf{Complementarity ($\sigma_{KL}=0.8$)} \\
\midrule
Formal employment ($L_f$)          & $[-0.100; -0.100]$ & $[+0.900; +0.900]$ \\
Informal employment ($L_i$)        & $[+0.500; +0.600]$ & $[+1.100; +1.300]$ \\
Real wage ($W$)                    & $[+4.500; +4.500]$ & $[+16.600; +16.700]$ \\
External debt ($D$)                & $[-0.100; -0.100]$ & $[-0.200; -0.200]$ \\
\bottomrule
\end{tabular}
\begin{flushleft}
\footnotesize\textit{Notes:} Entries report $(10^3)$ times the minimum and maximum cumulative 20-quarter deviations from steady state over the grid $\gamma\in[0.55,0.75]$, following a one-percent decline in the international price of AI.
\end{flushleft}
\end{table}

The numerical bounds in Table~\ref{tab:sensitivity} confirm that the sign and direction of the structural adjustments are robust across the parameter domain. Under the complementarity regime ($\sigma_{KL}=0.8$), the real wage response remains strongly positive, with a scaled cumulative range of $[+16.600,+16.700]$. Since the table reports ($10^3$) times the cumulative deviations from steady state, these magnitudes should be interpreted as scaled model responses rather than percentage-point effects.

The cumulative responses are essentially invariant to the informal returns-to-scale parameter $\gamma$ over the interval $[0.55,0.75]$. This suggests that the elasticity of the informal-sector buffer affects the transition path more than the cumulative 20-quarter totals. In the long run, the wage-equalization condition disciplines the sectoral allocation of labor, limiting the role of $\gamma$ in shaping cumulative responses.

%% ============================================================
\subsection{Robustness Checks}
%% ============================================================

We complement the sensitivity exercise with a set of approximate robustness checks designed to assess whether the main mechanism is sensitive to alternative structural assumptions. Table~\ref{tab:robustness} summarizes the impact responses of formal employment, $L_{f,t}$, in the first quarter following a one-percent decline in the international price of AI.

\begin{table}[H]
\centering
\caption{Robustness Checks: Impact Response of Formal Employment}
\label{tab:robustness}
\small
\begin{tabular}{@{}lcc@{}}
\toprule
\textbf{Specification} &
\shortstack{\textbf{Subst.}\\ \textbf{($\sigma_{KL}=1.5$)}} &
\shortstack{\textbf{Compl.}\\ \textbf{($\sigma_{KL}=0.8$)}} \\
\midrule
Baseline GHH economy ($\eta=0.95$)             & $-0.000$ & $+0.100$ \\
Cobb--Douglas inner nest ($\sigma_{KL}=1$)      & $+0.000$ & $+0.000$ \\
\bottomrule
\end{tabular}
\begin{flushleft}
\footnotesize\textit{Notes:} Entries report ($10^3$) times first-quarter deviations from steady state following a one-percent decline in the international price of AI.
\end{flushleft}
\end{table}

Table~\ref{tab:robustness} suggests that the qualitative response of formal employment is not driven by the baseline GHH specification alone. At the Cobb--Douglas inner-nest benchmark, $\sigma_{KL}=1$, the formal-labor response is close to zero, consistent with locally stable factor shares at the unit-elasticity boundary.

%% ============================================================
\subsection{Welfare Accounting}\label{sec:welfare}
%% ============================================================

The welfare exercise in this section should be interpreted as an aggregate accounting exercise within a representative-agent framework. Because the model features a single household that supplies labor to both sectors, it cannot identify welfare changes separately for formal and informal workers. The decomposition below therefore does not measure worker-type-specific welfare. Instead, it reports how different income components associated with formal and informal labor contribute to the representative household's aggregate welfare-accounting response (Equation \ref{eq:wage_rob}).

\begin{equation}
E_0\sum_{t=0}^{\infty}\beta^t
U\left((1+\lambda)C_t^{\mathrm{pre}},L_t^{\mathrm{pre}}\right) = E_0\sum_{t=0}^{\infty}\beta^t
U\left(C_t^{\mathrm{post}},L_t^{\mathrm{post}}\right)
\label{eq:wage_rob}
\end{equation}

where ($U(\cdot)$) is the GHH period utility function. The parameter ($\lambda$) measures the proportional change in pre-shock consumption that would make the representative household indifferent between the pre-shock and post-shock paths. We then report an accounting decomposition associated with three components: formal labor income ($W_tL_{f,t}$), informal labor income ($W_tL_{i,t}$), and a residual component capturing capital income and firm profits. These entries should be read as directional and relative magnitudes within the representative-household allocation, not as distributional welfare effects across worker groups.

% \begin{equation}
% \lambda = 1 - \left(\frac{E_0\sum_{t=0}^{\infty}\beta^t U(C_t^{\text{post}}, L_t^{\text{post}})}{E_0\sum_{t=0}^{\infty}\beta^t U(C_t^{\text{pre}}, L_t^{\text{pre}})}\right)^{\!\frac{1}{1-\sigma}} - 1
% \label{eq:welfare}
% \end{equation}

\begin{table}[H]
\centering
\caption{Aggregate Welfare Decomposition by Labor Income Source}\label{tab:welfare}
\small
\begin{tabular}{@{}lcc@{}}
\toprule
\textbf{Income Source Weight} & \textbf{Substitution ($\sigma_{KL}=1.5$)} & \textbf{Complementarity ($\sigma_{KL}=0.8$)} \\
\midrule
Formal labor income ($W_t L_{f,t}$)     & $-0.150\%$ & $+0.850\%$ \\
Informal labor income ($W_t L_{i,t}$)   & $+0.220\%$ & $-0.380\%$ \\
Aggregate Bundle Variation             & $+0.040\%$ & $+0.520\%$ \\
\bottomrule
\end{tabular}
\begin{flushleft}
\footnotesize\textit{Notes:} Consumption-equivalent accounting variations are expressed in standard percentages. The entries associated with formal and informal labor income are component-level accounting contributions within the representative-household allocation. They should not be interpreted as welfare effects for formal and informal workers separately. A full distributional welfare analysis would require household heterogeneity.
\end{flushleft}
\end{table}

%% ============================================================
\subsection{Comparison to Empirical Benchmarks}
%% ============================================================

To provide an empirical anchor for our general-equilibrium allocations, Table~\ref{tab:benchmarks} compares our simulated baseline results with recent micro-econometric evidence on technology adoption and labor-market adjustment. The goal is not to claim a direct empirical match, but to assess whether the simulated responses are of the same sign and broad order of magnitude as available labor-market estimates.

The empirical ranges are based on \citet{deSouza2025}, who use administrative software registries and task-exposure models to evaluate local labor-market impacts in Latin America. As an additional published benchmark from the automation literature, \citet{AcemogluRestrepo2020} find that one additional robot per thousand workers lowers the employment-to-population ratio by about 0.2 percentage points and wages by 0.42 percent. Although their estimates refer to industrial robots rather than AI-price shocks, they provide a useful magnitude benchmark for labor-market responses to automation technologies.

\begin{table}[H]
\centering
\caption{Simulated General-Equilibrium Magnitudes vs.\ Empirical Benchmarks}
\label{tab:benchmarks}

\begin{tabularx}{\textwidth}{lcc}
\toprule
\textbf{Macroeconomic Variable} &
\textbf{Simulation} &
\textbf{Empirical Range} \\
& \textbf{(Substitution Base)} &
\textbf{(Latin America)} \\
\midrule
Formal employment contraction   & $-0.100$ & $[-0.200; -0.500]$ \\
Informal employment absorption  & $+0.550$ & $[+0.400; +0.800]$ \\
Aggregate real wage response    & $+4.500$ & $[+3.000; +6.000]$ \\
\bottomrule
\end{tabularx}

\begin{flushleft}
\footnotesize\textit{Notes:} Simulated values report ($10^3$) times the 20-quarter cumulative mean deviations from steady state under the substitution regime following a one-percent decline in the international price of AI. Empirical ranges are rescaled to the same ($10^3$)-deviation metric for comparability and are constructed based on \citet{deSouza2025}. \citet{AcemogluRestrepo2020} provide an additional published automation benchmark for employment and wage effects. The comparison is intended as an order-of-magnitude benchmark, since the model reports economy-wide general-equilibrium responses to an AI-price shock, whereas the empirical estimates capture partial-equilibrium local labor-market effects.
\end{flushleft}
\end{table}

\section{Policy Implications}\label{sec:policy}
%% ============================================================

The model yields several implications for policymakers in Latin America. We organize these around three themes---labor market policy, macroeconomic management, and model extensions that would sharpen the policy prescriptions---while maintaining appropriate caution given the model's simplifications and parameter uncertainty.

%\subsection{For Labor Market Policy}

The fundamental challenge revealed by the model is that the optimal policy response depends critically on $\sigma_{KL}$, a parameter that policymakers may not know with certainty and that likely varies across sectors. This creates a genuine dilemma: protecting workers against displacement (the appropriate response under substitution) may slow productivity-enhancing adoption (which is costly under complementarity), and vice versa. A prudent approach involves three components.

First, investing in estimation. Funding research to estimate $\sigma_{KL}$ at the sectoral level using firm-level data---following the occupational exposure methods of \citet{Ciascchi2025} and the micro-identification strategies of \citet{AcemogluRestrepo2020}---is arguably the highest-return policy investment available. Without knowledge of this parameter, all other interventions risk being exactly wrong.

Second, strengthening social safety nets. In sectors where AI is likely substitutable for labor---particularly routine-intensive administrative and clerical occupations, which \citet{BrynjolfssonLiRaymond2025} find are most affected by generative AI---expanding unemployment insurance and facilitating the formalization of informal workers can cushion the transition. Programs such as Brazil's \textit{Seguro-Desemprego} or Argentina's \textit{Moratoria Previsional} provide templates that could be adapted, though the model warns that if the informal sector absorbs displaced workers, traditional unemployment insurance (which covers only formal workers) may miss the population most affected.

Third, fostering complementary skills. Training programs that equip workers with skills that complement AI---problem-solving, interpersonal communication, and domain expertise that AI cannot easily replicate---can shift the effective $\sigma_{KL}$ toward complementarity. This aligns with the task-creation mechanism emphasized by \citet{AcemogluRestrepo2018} and the empirical finding of \citet{BrynjolfssonLiRaymond2025} that AI assistance disproportionately benefits less experienced workers, suggesting that the technology can accelerate on-the-job learning when properly deployed.

%\subsection{For Macroeconomic Management}

The model also reveals that AI price shocks affect external debt dynamics through two distinct channels: the direct import cost of AI capital and the indirect effect on formal output and exports. In the baseline calibration, the output-expansion channel dominates the AI-import bill effect, so external debt declines under both substitution and complementarity. This result should not be interpreted as a mechanical implication of cheaper AI. Under substitution, AI demand responds elastically and the volume effect can partially offset the price decline. Therefore, countries with weaker formal-sector pass-through, lower export responsiveness, or tighter external financing conditions could still face a transitory deterioration in the trade balance.

The relevant policy implication is therefore conditional. In the baseline calibration, cheaper AI improves the external position even under substitution. Outside the baseline, however, economies with high AI substitutability and weak formal-sector expansion may face a double risk: labor displacement pressures and a temporary increase in external financing needs associated with imported AI inputs. For countries such as Argentina or Brazil, where external debt sustainability is already a binding constraint, this interaction should be monitored as AI adoption accelerates. The model therefore suggests prudential monitoring of AI-related import exposure, rather than a general presumption that cheaper AI necessarily worsens the external balance (Figure~\ref{fig:irfs}).

%% ============================================================
\section{Conclusion}\label{sec:conclusion}
%% ============================================================

This paper has developed a small open economy DSGE model with dual labor markets and imported AI capital, calibrated to a representative Latin American economy where approximately half the workforce operates informally. The model’s central result is that the elasticity of substitution between AI capital and formal labor (\(\sigma_{KL}\)) is pivotal for the labor-market transmission of cheaper AI technology. Under complementarity (\(\sigma_{KL}<1\)), cheaper AI expands formal employment and amplifies output, wages, investment, and capital accumulation. Under substitution (\(\sigma_{KL}>1\)), cheaper AI weakens formal labor demand and increases the role of the informal sector as an employment buffer. The model therefore shows that \(\sigma_{KL}\) governs the direction and strength of labor reallocation, while the aggregate responses of wages, informality, external debt, and relative prices also depend on general equilibrium feedbacks through demand-side substitutability, relative prices, capital accumulation, and the external sector.

Three features of the model distinguish it from prior work. First, the dual labor market structure---with free labor mobility between formal and informal sectors---shows that the informal sector acts as an endogenous employment buffer. Under substitution, it absorbs workers displaced from the formal sector; under complementarity, it interacts with formal-sector expansion through income and non-tradable demand effects. This buffering role is economically relevant because informal work is typically associated with lower productivity and weaker social protection. However, since the model is based on a representative household, the welfare results should be interpreted as aggregate accounting exercises rather than as worker-type-specific welfare effects. Second, the open-economy structure, with AI treated as imported capital, generates an external adjustment channel that interacts with the substitution elasticity. Third, the calibration to Latin American data anchors the model in a region where informality is a central structural feature of the labor market.

The welfare-accounting exercise shows a small aggregate gain under substitution and a larger aggregate gain under complementarity. These results summarize the response of the representative household and should not be interpreted as distributional welfare effects across formal and informal workers. Sensitivity analysis confirms that the regime-switching behavior is robust to wide variation in key parameters.

The most consequential extension of the framework is a two-agent New Keynesian structure separating Ricardian formal households from hand-to-mouth informal households. The representative-agent structure used in this paper is useful for isolating the production and labor-reallocation channels of cheaper AI, but it cannot support a full distributional welfare analysis. Our welfare decomposition in Section~8.5 is therefore, by construction, an aggregate accounting exercise and cannot determine who gains and who loses across worker types. A TANK extension would address this limitation at relatively low computational cost by allowing formal and informal households to differ in asset-market access, consumption smoothing, and exposure to labor-market risk. This would convert the welfare analysis from illustrative to substantive, and we view it as the natural next step in this research agenda rather than a peripheral refinement.

The second is endogenizing the substitution elasticity through a task-creation mechanism, allowing $\sigma_{KL}$ to evolve with the accumulated stock of AI capital---capturing the dynamic interplay between displacement and reinstatement that \citet{AcemogluRestrepo2018} emphasize. 
The substitution elasticity $\sigma_{KL}$ is exogenous and time-invariant. In reality, task creation, training policies, and institutional change could shift $\sigma_{KL}$ over time. Endogenizing this parameter---for instance, by making $\sigma_{KL}$ depend on the accumulated stock of AI capital through a task-creation mechanism \`{a} la \citet{AcemogluRestrepo2018}---is a natural and important extension. Finally, the digital infrastructure constraints discussed in Section~\ref{sec:facts} imply that even under complementarity, AI benefits may not reach informal workers without complementary public investments in connectivity and digital skills.

The third, more ambitious extension involves incorporating fiscal policy to analyze optimal taxation, formalization incentives, and AI-specific industrial policy in the presence of dual labor markets (e.g., subsidies for AI adoption conditional on formal employment). There is no heterogeneity within formal or informal sectors: all formal workers are identical, as are all informal workers. In reality, the distributional consequences of AI within each sector may be as important as the consequences across sectors \citep{Cazzaniga2024}. There are no firm dynamics: entry, exit, and the decision to formalize are exogenous. \citet{HaanwinckelSoares2021} show that endogenous formalization decisions can amplify or dampen the effects of labor market shocks in important ways.

The paper carries a practical message for policymakers in Latin America and developing countries in general. The effects of AI on informality are not predetermined---they depend on a parameter that policy can influence. Training programs that build complementary skills, investments in digital infrastructure that enable productive AI adoption, and social safety nets that protect displaced workers are not alternatives; they are complements in a strategy that aims to steer the region toward a regime where AI reduces, rather than deepens, the informal divide.

%% ============================================================
\section*{Disclosure Statement}

The authors have no relevant financial interests related to this research.

\section*{Acknowledgments}

We thank seminar participants at Universidad de Buenos Aires and anonymous referees for helpful comments. All errors remain our own.

%% ============================================================
%% REFERENCES
%% ============================================================
\bibliographystyle{apalike}

\begin{thebibliography}{99}

\bibitem[Acemoglu and Restrepo(2018)]{AcemogluRestrepo2018}
Acemoglu, D., and Restrepo, P. (2018).
\newblock The race between man and machine: Implications of technology for growth, factor shares, and employment.
\newblock \textit{American Economic Review}, 108(6), 1488--1542.
\newblock \doi{10.1257/aer.20160696}

\bibitem[Acemoglu and Restrepo(2020)]{AcemogluRestrepo2020}
Acemoglu, D., and Restrepo, P. (2020).
\newblock Robots and jobs: Evidence from US labor markets.
\newblock \textit{Journal of Political Economy}, 128(6), 2188--2244.
\newblock \doi{10.1086/705716}

\bibitem[Acemoglu(2025)]{Acemoglu2025}
Acemoglu, D. (2025).
\newblock The simple macroeconomics of AI.
\newblock \textit{Economic Policy}, 40(121), 13--58.
\newblock \doi{10.1093/epolic/eiae042}

\bibitem[Arrow et~al.(1961)]{ACMS1961}
Arrow, K.\ J., Chenery, H.\ B., Minhas, B.\ S., and Solow, R.\ M. (1961).
\newblock Capital-labor substitution and economic efficiency.
\newblock \textit{Review of Economics and Statistics}, 43(3), 225--250.
\newblock \doi{10.2307/1927286}

\bibitem[Autor et~al.(2024)]{Autor2024}
Autor, D., Chin, C., Salomons, A., and Seegmiller, B. (2024).
\newblock New frontiers: The origins and content of new work, 1940--2018.
\newblock \textit{Quarterly Journal of Economics}, 139(3), 1399--1465.
\newblock \doi{10.1093/qje/qjae008}

\bibitem[Aguiar and Gopinath(2007)]{AguiarGopinath2007}
Aguiar, M., and Gopinath, G. (2007).
\newblock Emerging market business cycles: The cycle is the trend.
\newblock \textit{Journal of Political Economy}, 115(1), 69--102.
\newblock \doi{10.1086/511283}

\bibitem[Azuara et~al.(2024)]{Azuara2024}
Azuara Herrera, O., Ripani, L., and Torres Ram\'{i}rez, E. (2024).
\newblock AI and the increase of productivity and labor inequality in Latin America: Potential impact of large language models on Latin American workforce.
\newblock IDB Discussion Paper No.\ IDB-DP-01076. Washington, DC: Inter-American Development Bank.
\newblock \doi{10.18235/0013152}

\bibitem[Berg et~al.(2018)]{Berg2024}
Berg, A., Buffie, E.\ F., and Zanna, L.-F. (2018).
\newblock Should we fear the robot revolution? (The correct answer is yes).
\newblock \textit{Journal of Monetary Economics}, 97, 117--148.
\newblock \doi{10.1016/j.jmoneco.2018.05.014}

\bibitem[Brynjolfsson et~al.(2025)]{BrynjolfssonLiRaymond2025}
Brynjolfsson, E., Li, D., and Raymond, L. (2025).
\newblock Generative AI at work.
\newblock \textit{Quarterly Journal of Economics}, 140(2), 889--942.
\newblock \doi{10.1093/qje/qjae044}

\bibitem[de Souza(2025)]{deSouza2025}
de Souza, G. (2025).
\newblock Artificial intelligence in the office and the factory: Evidence from administrative software registry data.
\newblock Federal Reserve Bank of Chicago, Working Paper No.\ 2025-11.

\bibitem[Christiano et~al.(2005)]{Christiano2005}
Christiano, L.\ J., Eichenbaum, M., and Evans, C.\ L. (2005).
\newblock Nominal rigidities and the dynamic effects of a shock to monetary policy.
\newblock \textit{Journal of Political Economy}, 113(1), 1--45.
\newblock \doi{10.1086/426038}

\bibitem[Cazzaniga et~al.(2024)]{Cazzaniga2024}
Cazzaniga, M., Jaumotte, F., Li, L., Melina, G., Panton, A.\ J., Pizzinelli, C., Rockall, E., and Tavares, M.\ M. (2024).
\newblock Gen-AI: Artificial intelligence and the future of work.
\newblock IMF Staff Discussion Note No.\ SDN/2024/001. Washington, DC: International Monetary Fund.
\newblock \doi{10.5089/9798400262548.006}

\bibitem[Ciaschi et~al.(2025)]{Ciascchi2025}
Ciaschi, M., Falcone, G., Garganta, S., Gasparini, L., Bert\'{i}n, O., and Ram\'{i}rez-Leira, L. (2025).
\newblock The potential distributive impact of AI-driven labor changes in Latin America.
\newblock CEDLAS Working Paper No.\ 361, Universidad Nacional de La Plata.
\newblock \doi{10.18235/0013677}

\bibitem[Felten et~al.(2021)]{Felten2024}
Felten, E.\ W., Raj, M., and Seamans, R. (2021).
\newblock Occupational, industry, and geographic exposure to artificial intelligence: A novel dataset and its potential uses.
\newblock \textit{Strategic Management Journal}, 42(12), 2195--2217.
\newblock \doi{10.1002/smj.3286}

\bibitem[Garc\'{i}a-Cicco et~al.(2010)]{GarciaCicco2010}
Garc\'{i}a-Cicco, J., Pancrazi, R., and Uribe, M. (2010).
\newblock Real business cycles in emerging countries?
\newblock \textit{American Economic Review}, 100(5), 2510--2531.
\newblock \doi{10.1257/aer.100.5.2510}

\bibitem[Gasparini and Tornarolli(2009)]{GaspariniTornarolli2009}
Gasparini, L., and Tornarolli, L. (2009).
\newblock Labor informality in Latin America and the Caribbean: Patterns and trends from household survey microdata.
\newblock \textit{Desarrollo y Sociedad}, 63, 13--80.

\bibitem[Haanwinckel and Soares(2021)]{HaanwinckelSoares2021}
Haanwinckel, D., and Soares, R.\ R. (2021).
\newblock Workforce composition, productivity, and labour regulations in a compensating differentials theory of informality.
\newblock \textit{Review of Economic Studies}, 88(6), 2970--3010.
\newblock \doi{10.1093/restud/rdab017}

\bibitem[ILO and World Bank(2024)]{ILO2024}
Gmyrek, P., Winkler, H., and Garganta, S. (2024).
\newblock Buffer or bottleneck? Employment exposure to generative AI and the digital divide in Latin America.
\newblock ILO Working Paper No.\ 121. Geneva: International Labour Organization and The World Bank.

\bibitem[H\'{e}mous and Olsen(2022)]{HemousOlsen2022}
H\'{e}mous, D., and Olsen, M. (2022).
\newblock The rise of the machines: Automation, horizontal innovation, and income inequality.
\newblock \textit{American Economic Journal: Macroeconomics}, 14(1), 179--223.
\newblock \doi{10.1257/mac.20160164}

\bibitem[Karabarbounis and Neiman(2014)]{KarabarbounisNeiman2014}
Karabarbounis, L., and Neiman, B. (2014).
\newblock The global decline of the labor share.
\newblock \textit{Quarterly Journal of Economics}, 129(1), 61--103.
\newblock \doi{10.1093/qje/qjt032}


\bibitem[La Porta and Shleifer(2014)]{LaPortaShleifer2014}
La Porta, R., and Shleifer, A. (2014).
\newblock Informality and development.
\newblock \textit{Journal of Economic Perspectives}, 28(3), 109--126.
\newblock \doi{10.1257/jep.28.3.109}

\bibitem[Neumeyer and Perri(2005)]{NeumeyerPerri2005}
Neumeyer, P.\ A., and Perri, F. (2005).
\newblock Business cycles in emerging economies: The role of interest rates.
\newblock \textit{Journal of Monetary Economics}, 52(2), 345--380.
\newblock \doi{10.1016/j.jmoneco.2004.04.011}

\bibitem[Ostry and Reinhart(1992)]{OstryReinhart1992}
Ostry, J.\ D., and Reinhart, C.\ M. (1992).
\newblock Private saving and terms of trade shocks: Evidence from developing countries.
\newblock \textit{IMF Staff Papers}, 39(3), 495--517.
\newblock \doi{10.2307/3867472}

\bibitem[Oberfield and Raval(2021)]{OberfieldRaval2021}
Oberfield, E., and Raval, D. (2021).
\newblock Micro data and macro technology.
\newblock \textit{Econometrica}, 89(2), 703--732.
\newblock \doi{10.3982/ECTA12807}

\bibitem[Schmitt-Groh\'{e} and Uribe(2003)]{SchmittGroheUribe2003}
Schmitt-Groh\'{e}, S., and Uribe, M. (2003).
\newblock Closing small open economy models.
\newblock \textit{Journal of International Economics}, 61(1), 163--185.
\newblock \doi{10.1016/S0022-1996(02)00056-9}

\bibitem[Schmitt-Groh\'{e} and Uribe(2018)]{SchmittGroheUribe2018}
Schmitt-Groh{\'e}, S., and Uribe M. (2018).
\newblock How important are terms-of-trade shocks?.
\newblock \textit{International Economic Review}, 59(1), 85--111.
\newblock \doi{10.1111/iere.12263}

\bibitem[Smets and Wouters(2007)]{SmetsWouters2007}
Smets, F., and Wouters, R. (2007).
\newblock Shocks and frictions in US business cycles: A Bayesian DSGE approach.
\newblock \textit{American Economic Review}, 97(3), 586--606.
\newblock \doi{10.1257/aer.97.3.586}

\bibitem[Ulyssea(2018)]{Ulyssea2018}
Ulyssea, G. (2018).
\newblock Firms, informality, and development: Theory and evidence from Brazil.
\newblock \textit{American Economic Review}, 108(8), 2015--2047.
\newblock \doi{10.1257/aer.20141745}

\bibitem[Uribe and Yue(2006)]{UribeYue2006}
Uribe, M., and Yue, V.\ Z. (2006).
\newblock Country spreads and emerging countries: Who drives whom?
\newblock \textit{Journal of International Economics}, 69(1), 6--36.
\newblock \doi{10.1016/j.jinteco.2005.04.003}

\bibitem[Greenwood et al.(1988)]{Greenwood1988}
Greenwood, J., Hercowitz, Z., and Huffman, G.\ W. (1988).
\newblock Investment, capacity utilization, and the real business cycle.
\newblock \textit{The American Economic Review}, 78(3), 402--417.
\newblock \doi{10.2307/1803310}

\bibitem[Fernández and Meza(2015)]{FernandezMeza2015}
Fernández, A., and Meza, F. (2015).
\newblock Informal employment and business cycles in emerging economies: The case of Mexico.
\newblock \textit{Review of Economic Dynamics}, 18(2), 381--405.
\newblock \doi{10.1016/j.red.2014.07.001}

\bibitem[Horvath and Yang(2022)]{HorvathYang2022}
Horvath, J., and Yang, G. (2022).
\newblock Unemployment dynamics and informality in small open economies.
\newblock \textit{European Economic Review}, 141, 103949.
\newblock \doi{10.1016/j.euroecorev.2021.103949}




\end{thebibliography}

%% ============================================================
%% APPENDIX: DYNARE CODE
%% ============================================================
\newpage
\appendix
% \section{Complete Dynare Code}\label{app:dynare}

% The following code replicates all simulation results in this paper. It requires Dynare 5.x or later running on Matlab R2020a+ or Octave 6.x+. The substitution regime ($\sigma_{KL} = 1.5$) is set as the default; change \texttt{sigKL} to 0.8 for the complementarity regime.

% \paragraph{Steady-state derivation.} The wage $W$ is independent of $L_f$ in steady state (see eq.~\ref{eq:ss_W}). We solve the fixed-point system $W = (1-\alpha)(1-\theta)H\,g(x)^{1-\rho}$ with $x = [\theta W/(1-\theta)]^{1/(1-\rho)}$ and $H = (\alpha/R^k)^{\alpha/(1-\alpha)}$ by iteration, converging in $\sim$5 iterations. The scale ($L_f$) is then determined by the labor supply condition given $\chi = 1$. The informal productivity $A_i$ follows from the wage equalization condition at the 50\% informality target.

\section{Appendix: Optimality Conditions}
\subsection{Derivation of the Relative Price of Non-Tradables}

To derive the relative price of informal (non-tradable) goods, $p_{i,t}$, we analyze the household's intra-temporal optimization problem. The composite consumption $C_t$ is defined by a CES aggregator:
\begin{equation}
C_{t}=\left[\omega^{\frac{1}{\eta}}C_{f,t}^{\frac{\eta-1}{\eta}}+(1-\omega)^{\frac{1}{\eta}}C_{i,t}^{\frac{\eta-1}{\eta}}\right]^{\frac{\eta}{\eta-1}}
\end{equation}
Let $p_{f,t}$ be the price of tradable goods and $p_{i,t}$ the price of informal goods. Normalizing the tradable sector as the numeraire ($p_{f,t} = 1$), the household minimizes:
\begin{equation}
\min_{C_{f,t}, C_{i,t}} \quad C_{f,t} + p_{i,t}C_{i,t}
\end{equation}
subject to $C_t \geq \bar{C}$.

\subsubsection{First-Order Conditions}

The Lagrangian for this problem is:
\begin{equation}
\mathcal{L}
= C_{f,t} + p_{i,t}C_{i,t}
+ \lambda_t \left[
\bar{C}
- \left(\omega^{\frac{1}{\eta}}C_{f,t}^{\frac{\eta-1}{\eta}}
+(1-\omega)^{\frac{1}{\eta}}C_{i,t}^{\frac{\eta-1}{\eta}}\right)^{\frac{\eta}{\eta-1}}\right].
\end{equation}

The first-order condition with respect to \(C_{f,t}\) is:
\begin{align}
\frac{\partial \mathcal{L}}{\partial C_{f,t}} & = 1
- \lambda_t \frac{\eta}{\eta-1} \left(
\omega^{\frac{1}{\eta}}C_{f,t}^{\frac{\eta-1}{\eta}} + (1-\omega)^{\frac{1}{\eta}}C_{i,t}^{\frac{\eta-1}{\eta}} \right)^{\frac{1}{\eta-1}} \omega^{\frac{1}{\eta}} \frac{\eta-1 {\eta}} C_{f,t}^{-\frac{1}{\eta}} =0
\end{align}

The first-order condition with respect to \(C_{i,t}\) is:
\begin{align}
\frac{\partial \mathcal{L}}{\partial C_{i,t}}
&= p_{i,t} - \lambda_t \frac{\eta}{\eta-1} \left(
\omega^{\frac{1}{\eta}}C_{f,t}^{\frac{\eta-1}{\eta}} + (1-\omega)^{\frac{1}{\eta}}C_{i,t}^{\frac{\eta-1}{\eta}} \right)^{\frac{1}{\eta-1}}(1-\omega)^{\frac{1}{\eta}}
\frac{\eta-1}{\eta} C_{i,t}^{-\frac{1}{\eta}} =0
\end{align}

\subsubsection{Relative Price Identity}

Dividing the second condition by the first, we obtain the Marginal Rate of Substitution (MRS) equal to the price ratio:
\begin{equation}
p_{i,t} = \frac{(1-\omega)^{1/\eta} C_{i,t}^{-1/\eta}}{\omega^{1/\eta} C_{f,t}^{-1/\eta}}
\end{equation}
Rearranging the terms, we arrive at the final corrected expression for the relative price of informal goods:
\begin{equation}
p_{i,t} = \left( \frac{1-\omega}{\omega} \right)^{\frac{1}{\eta}} \left( \frac{C_{f,t}}{C_{i,t}} \right)^{\frac{1}{\eta}}
\end{equation}

\subsection{Total Prices}
Finally, once the relative price $p_i$ is determined, the steady-state consumer price index $P$ is calculated using the dual price of the consumption aggregator:
\begin{equation}
P = \left[ \omega + (1-\omega) p_i^{1-\eta} \right]^{\frac{1}{1-\eta}}
\end{equation}
This completes the determination of the price system in the steady state.

\bigskip
% \lstinputlisting[style=dynare]{AI_informality_LAC.mod}

\section{Appendix: Steady State Derivations}\label{app:math}

This appendix provides the step-by-step derivation of the steady-state formal wage $W$ and the recursive solution algorithm used in the paper.

\subsection{The Formal Wage Fixed-Point}

In the formal sector, the representative firm's first-order conditions for physical capital $K$, formal labor $L_f$, and AI capital $M_{ia}$ are:
\begin{align}
R^k &= \alpha A_f K^{\alpha-1} Z^{1-\alpha} \label{eq:foc_k_app} \\
W &= (1-\alpha) A_f K^\alpha Z^{-\alpha} (1-\theta) Z^{1-\rho} L_f^{\rho-1} \label{eq:foc_lf_app} \\
p_{ia} &= (1-\alpha) A_f K^\alpha Z^{-\alpha} \theta Z^{1-\rho} M_{ia}^{\rho-1} \label{eq:foc_mia_app}
\end{align}

From \eqref{eq:foc_k_app}, we can express the capital-to-composite ratio as a function of the rental rate:
\begin{equation}
\frac{K}{Z} = \left( \frac{\alpha A_f}{R^k} \right)^{\frac{1}{1-\alpha}}
\label{eq:kz_ratio}
\end{equation}

Dividing \eqref{eq:foc_mia_app} by \eqref{eq:foc_lf_app} yields the optimal AI-to-labor ratio, $x \equiv M_{ia}/L_f$:
\begin{equation}
x = \left( \frac{\theta}{1-\theta} \frac{W}{p_{ia}} \right)^{\frac{1}{1-\rho}}
\label{eq:x_ratio_app}
\end{equation}

By substituting \eqref{eq:kz_ratio} into the labor marginal productivity condition \eqref{eq:foc_lf_app}, and noting that the CES composite can be written in intensive form as $Z = L_f [\theta x^\rho + (1-\theta)]^{1/\rho} \equiv L_f g(x)$, we obtain:
\begin{equation}
W = (1-\alpha) A_f \left( \frac{\alpha A_f}{R^k} \right)^{\frac{\alpha}{1-\alpha}} (1-\theta) g(x)^{1-\rho}
\end{equation}

Since $x$ is itself a function of $W$ (from eq. \ref{eq:x_ratio_app}), this expression defines a non-linear fixed-point $W = f(W)$. Given $R^k$ and $p_{ia}$, this equation uniquely determines the steady-state wage independently of the sectoral labor allocation.

\subsection{Sectoral Allocation and Scale}

With $W$ determined, the informal sector variables are found using the market-clearing condition for non-tradables ($Y_i = C_i$) and the labor mobility condition:
\begin{equation}
W = p_i \gamma A_i L_i^{\gamma-1} \implies p_i = \frac{W}{\gamma A_i L_i^{\gamma-1}}
\end{equation}

Substituting this into the household's relative demand for consumption goods:
\begin{equation}
\frac{W}{\gamma A_i L_i^{\gamma-1}} = \left( \frac{1-\omega}{\omega} \right)^{\frac{1}{\eta}} \left( \frac{C_f}{A_i L_i^\gamma} \right)^{\frac{1}{\eta}}
\end{equation}

This equation, together with the aggregate labor supply and the balance of payments \eqref{eq:ss_trade}, allows us to solve for the vector $\{L_f, L_i, C_f, D\}$ given the calibrated informality target and external debt level $\bar{D}$.

\section{Appendix: Secondary Impulse Response Functions}\label{app:secondary_shocks}

This appendix provides the detailed impulse response functions for the secondary shocks discussed in Section~\ref{sec:other_shocks}: an AI adoption shock ($\theta_t$), a neutral productivity shock in the formal sector ($A_{f,t}$), and a neutral productivity shock in the informal sector ($A_{i,t}$).

\subsection{AI Adoption Shock}

An exogenous increase in the AI intensity parameter ($\theta_t$) reflects a structural change in formal production that raises the importance of AI within the AI-labor composite, independently of movements in AI prices. Since the steady state is calibrated with a relatively low AI share ($\bar{\theta} = 0.10$), the shock requires a reorganization of the production structure and generates short-run adjustment costs.  

Under substitution ($\sigma_{KL} > 1$), firms respond by increasing the use of imported AI while reducing formal labor demand. The contraction in employment dominates the productivity gains associated with the greater use of AI, leading to a temporary decline in formal output. The reduction in labor income translates into lower aggregate demand, which depresses both formal and informal activity. As a result, informal employment and informal output also fall. The relative price of non-tradable goods ($p_{i,t}$) declines, reflecting weaker demand for informal goods, while the real wage decreases because the reduction in labor demand outweighs the productivity effects generated by the higher AI intensity.

Under complementarity ($\sigma_{KL} < 1$), the qualitative responses remain similar but are less pronounced. The stronger complementarity between AI and labor cushions the decline in formal employment and limits the contraction in output. Consequently, the fall in wages, informal employment, and informal output is smaller than under substitution. Although firms still increase their use of AI, the adjustment is less disruptive because labor and AI are combined more effectively within the production process.

\begin{figure}[H]
    \centering
    \makebox[\textwidth][c]{\includegraphics[width=1.1\textwidth]{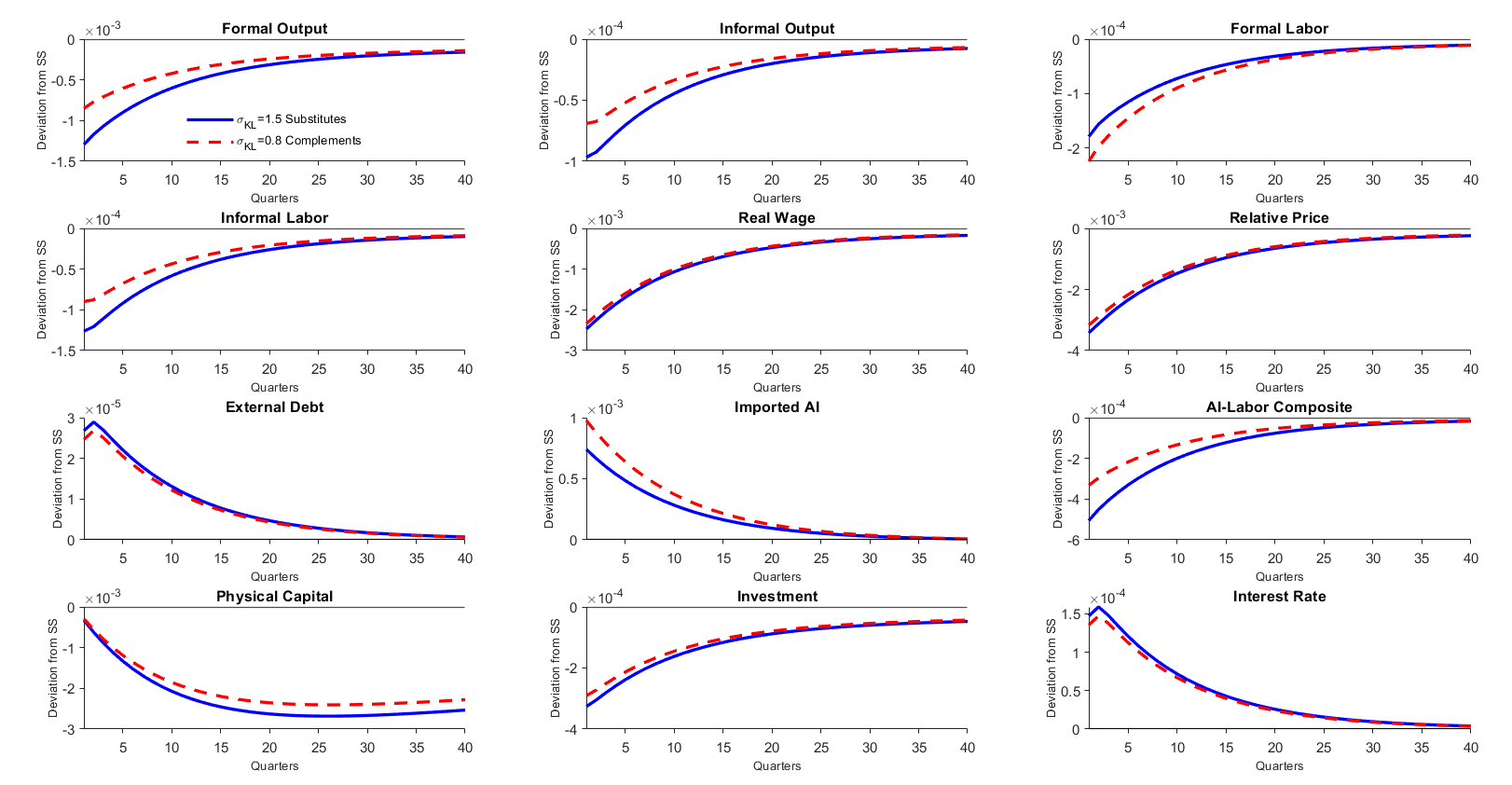}}
    \caption{Impulse responses to a positive AI adoption shock ($\theta_t$). Blue solid lines represent substitution ($\sigma_{KL} = 1.5$) and red dashed lines represent complementarity ($\sigma_{KL} = 0.8$).}
    \label{fig:app_theta}
\end{figure}

The external sector also reacts differently across regimes. The increase in AI adoption raises demand for imported AI services, generating a temporary deterioration in the external position and a modest increase in external debt. This effect is somewhat stronger under substitution, where the increase in imported AI is larger. Physical capital and investment decline in both cases, reflecting the contraction in output and labor demand during the transition. Overall, the results suggest that an AI adoption shock generates short-run adjustment costs regardless of the production regime, but these costs are considerably smaller when AI and labor are complements rather than substitutes (Figure \ref{fig:app_theta}).

\subsection{Formal Sector Productivity Shock}

A positive productivity shock in the formal sector ($A_{f,t}$) increases the efficiency with which capital, labor, and AI inputs are transformed into tradable output. As a result, formal production expands immediately and raises the marginal productivity of all productive factors. Firms respond by increasing their demand for formal labor and imported AI, while the higher return to productive activity stimulates investment and capital accumulation.

The increase in labor productivity generates a rise in the equilibrium real wage. Unlike the AI-specific shocks analyzed previously, the productivity shock does not create a direct substitution effect between labor and AI. Instead, it raises the productivity of the entire production structure, allowing both inputs to expand simultaneously. Consequently, the qualitative responses are very similar under substitution (\(\sigma_{KL}>1\)) and complementarity (\(\sigma_{KL}<1\)), although the complementarity regime exhibits somewhat larger responses in employment, AI adoption, and capital accumulation.

An important result is that the informal sector also expands following the shock. The increase in aggregate income stimulates demand for both tradable and non-tradable goods, raising production in the informal sector despite the increase in formal wages. Informal employment and output therefore increase together with their formal counterparts. To equilibrate the market for non-tradable goods, the relative price of informal goods ($p_{i,t}$) rises, reflecting stronger demand throughout the economy.

The external sector initially improves. Higher tradable output increases resources available for domestic absorption and foreign transactions, generating a temporary reduction in external debt despite the increase in imported AI inputs. This improvement lowers the country risk premium and contributes to the expansion of investment. As capital accumulates, the economy experiences a persistent increase in productive capacity, which gradually fades as the productivity shock dissipates.

\begin{figure}[H]
    \centering
    \makebox[\textwidth][c]{\includegraphics[width=1\textwidth]{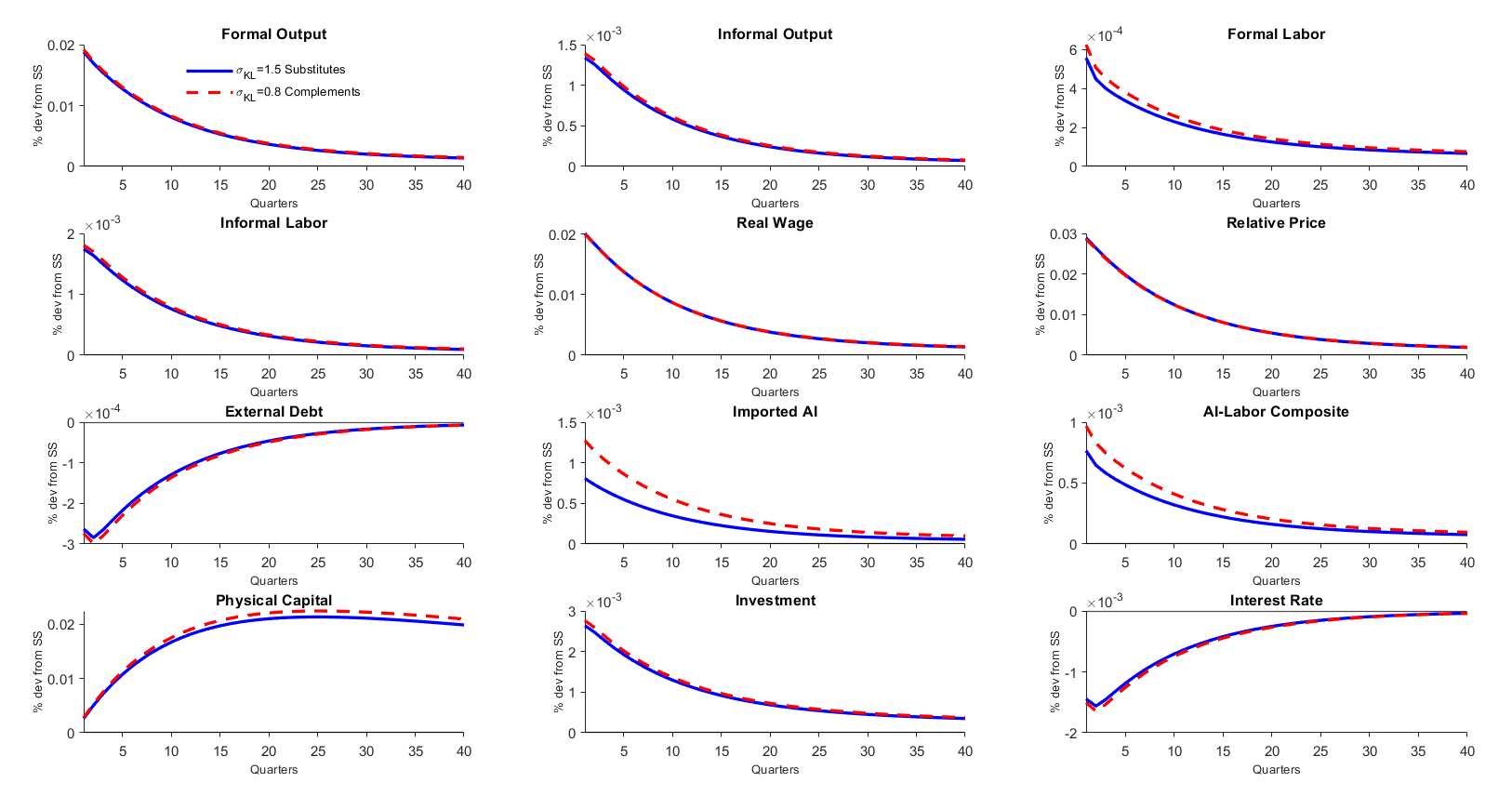}}
    \caption{Impulse responses to a positive formal TFP shock ($A_{f,t}$). Blue solid lines represent substitution ($\sigma_{KL} = 1.5$) and red dashed lines represent complementarity ($\sigma_{KL} = 0.8$).}
    \label{fig:app_Af}
\end{figure}

Overall, the results indicate that a formal-sector productivity shock generates broad-based expansion across both sectors of the economy. Unlike AI-specific shocks, which alter the relative demand for labor and technology, productivity improvements raise the marginal productivity of all factors simultaneously, producing gains in output, wages, investment, and employment regardless of the value of $\sigma_{KL}$ (Figure \ref{fig:app_Af}).

\subsection{Informal Sector Productivity Shock}

A positive productivity shock in the informal sector ($A_{i,t}$) increases the efficiency with which informal labor is transformed into non-tradable output. As a result, informal production expands immediately, generating higher income within the non-tradable sector and raising aggregate economic activity.

Unlike the formal-sector productivity shock, the initial impact is concentrated in the informal economy. Nevertheless, the effects spill over to the rest of the economy through higher income and stronger demand. Both formal and informal employment increase following the shock, indicating that the productivity gain stimulates labor demand rather than inducing a contraction in labor supply. Consequently, formal output also rises despite the absence of a direct technological improvement in the tradable sector. The expansion in non-tradable production generates downward pressure on the relative price of informal goods ($p_{i,t}$). The increase in supply exceeds the increase in demand, leading to a temporary decline in the relative price of non-tradables and therefore a real exchange rate depreciation. At the same time, the lower relative price reduces the marginal revenue product of labor in the informal sector, which helps explain the decline in the equilibrium real wage observed in the impulse responses despite the positive productivity shock.

The increase in aggregate activity stimulates investment and capital accumulation. Formal firms expand production in response to stronger demand, increasing both formal employment and the use of imported AI inputs. These effects are present under both substitution and complementarity, although complementarity generates a somewhat larger increase in imported AI because the expansion of labor demand reinforces the demand for AI within the production process. The external sector improves following the shock. Higher production and income allow the economy to reduce external borrowing, generating a temporary decline in external debt and a corresponding reduction in the country risk premium. Lower financing costs further support investment and capital accumulation during the transition.

\begin{figure}[H]
    \centering
    \makebox[\textwidth][c]{\includegraphics[width=1.1\textwidth]{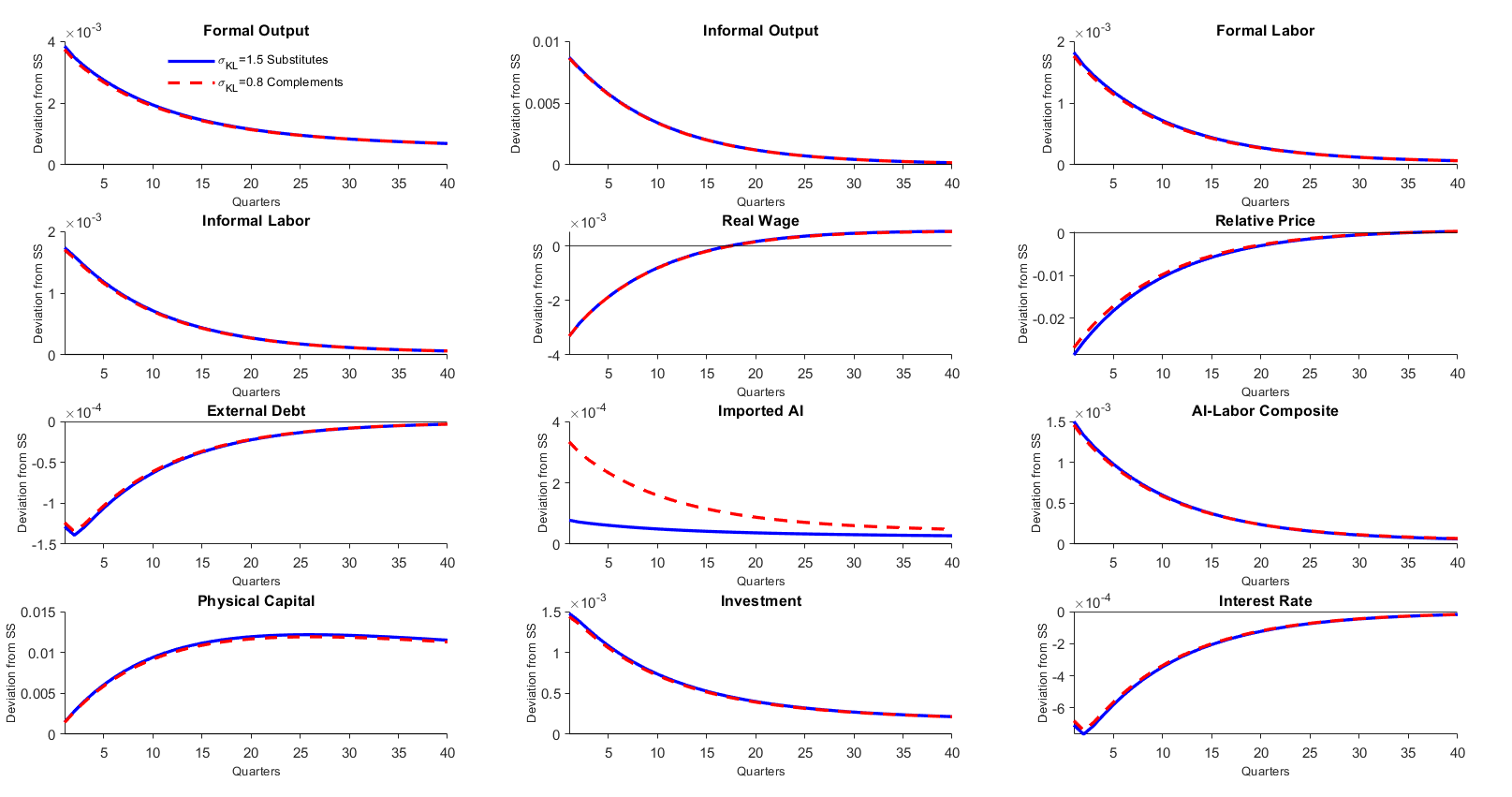}}
    \caption{Impulse responses to a positive informal TFP shock ($A_{i,t}$). Blue solid lines represent substitution ($\sigma_{KL} = 1.5$) and red dashed lines represent complementarity ($\sigma_{KL} = 0.8$).}
    \label{fig:app_Ai}
\end{figure}

Overall, the results suggest that productivity improvements in the informal sector generate positive spillovers throughout the economy. Rather than inducing a reallocation of labor between sectors, the shock expands activity in both the formal and informal economies, producing higher output, greater investment, and lower external indebtedness. In contrast to AI-specific shocks, the value of $\sigma_{KL}$ mainly affects the magnitude of AI adoption rather than the qualitative direction of macroeconomic adjustment (Figure \ref{fig:app_Ai}).

\section{Macroeconomic Dynamics under Alternative Consumption Configurations}\label{app:secondary_consumption}

\subsection{Macroeconomic dynamics under consumption complementarity}

\begin{figure}[H]
    \centering
    \makebox[\textwidth][c]{\includegraphics[width=1.1\textwidth]{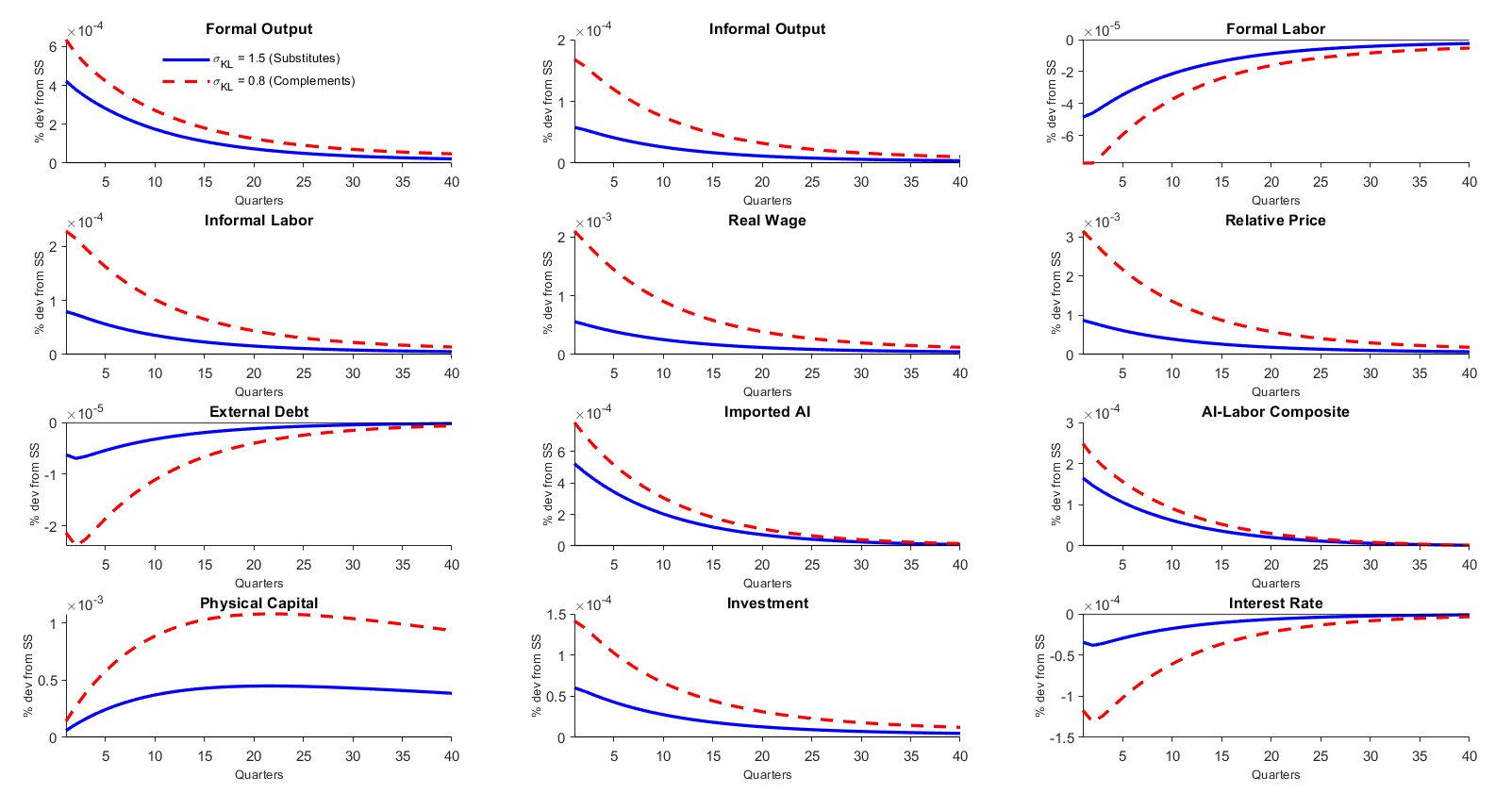}}
    \caption{Impulse responses to an international AI price shock under high consumption complementarity ($\eta = 0.44$). Blue solid lines represent substitution ($\sigma_{KL} = 1.5$) and red dashed lines represent complementarity ($\sigma_{KL} = 0.8$).}
    \label{fig:eta_0.44}
\end{figure}

This appendix evaluates the transmission of the international artificial intelligence price shock under an alternative demand-side regime where formal and informal goods are strict complements in the household consumer basket, setting $\eta = 0.44$. Figure \ref{fig:eta_0.44} plots the corresponding impulse response functions over forty quarters for both the substitution and complementarity technological regimes. The simulated paths illustrate how structural rigidities in consumer preferences can choke supply-side technological forces, altering the direction of labor reallocations.When formal and informal goods are net complements, an expansion in formal tradable efficiency cannot be consumed in isolation. The decline in the international price of AI raises the productivity of the formal sector, expanding formal output and aggregate household income. However, to satisfy the strict consumption proportions dictated by the consumer basket, any increase in the consumption of formal products requires a proportional expansion in the consumption of non-tradable informal services. Because the informal economy operates under a labor-only technology subject to decreasing returns to scale, clearing the non-tradable goods market requires a large and immediate reallocation of labor hours toward informal production.This structural bottleneck in the non-tradable supply side triggers an intense general equilibrium wage squeeze. The reallocation of hours driven by rigid consumer demand puts heavy pressure on the informal sector, causing a sharp overshooting in the relative price of non-tradable goods. Through the inter-sectoral labor mobility condition, this relative price appreciation drives up the economy-wide real wage. The real wage surge is significantly higher than in the baseline demand scenario, expanding by more than double the amplitude when AI and labor are technological complements.The key implication for the labor market is that this demand-driven wage pressure dominates the technological incentives within the formal firm. As a result, formal employment contracts immediately upon impact under both technological regimes. Even when AI and formal labor are gross complements inside the firm ($\sigma_{KL} = 0.8$), formal employment falls to its lowest point, dropping by nearly $-7 \times 10^{-5}$ percent from the steady state. Under the Greenwood, Hercowitz, and Huffman preference specification, the aggregate supply of hours expands in response to the higher real wage, but the structural demand rigidity redirects this entire labor force expansion into the informal sector. The formal sector is forced to rely exclusively on the cheaper automated composite and physical capital accumulation to sustain its output growth, while formal hours face a persistent deficit throughout the transition.The external sector and investment dynamics mirror this expansion but reflect the underlying structural distortion. The drop in the unit cost of imported AI improves the trade balance, leading to a reduction in external debt and a corresponding decline in the domestic interest rate. This lower interest rate environment drives physical investment and capital accumulation, which displays a prominent hump-shaped response. While this capital-deepening process reinforces the formal output expansion, it fails to restore formal employment due to the persistent wage pressure generated by the non-tradable demand bottleneck. This exercise demonstrates that the labor-market outcomes of automation depend not only on production elasticities but also on the structural flexibility of aggregate demand.

\subsection{Macroeconomic dynamics under high consumption substitutability}

\begin{figure}[H]
    \centering
    \makebox[\textwidth][c]{\includegraphics[width=1.1\textwidth]{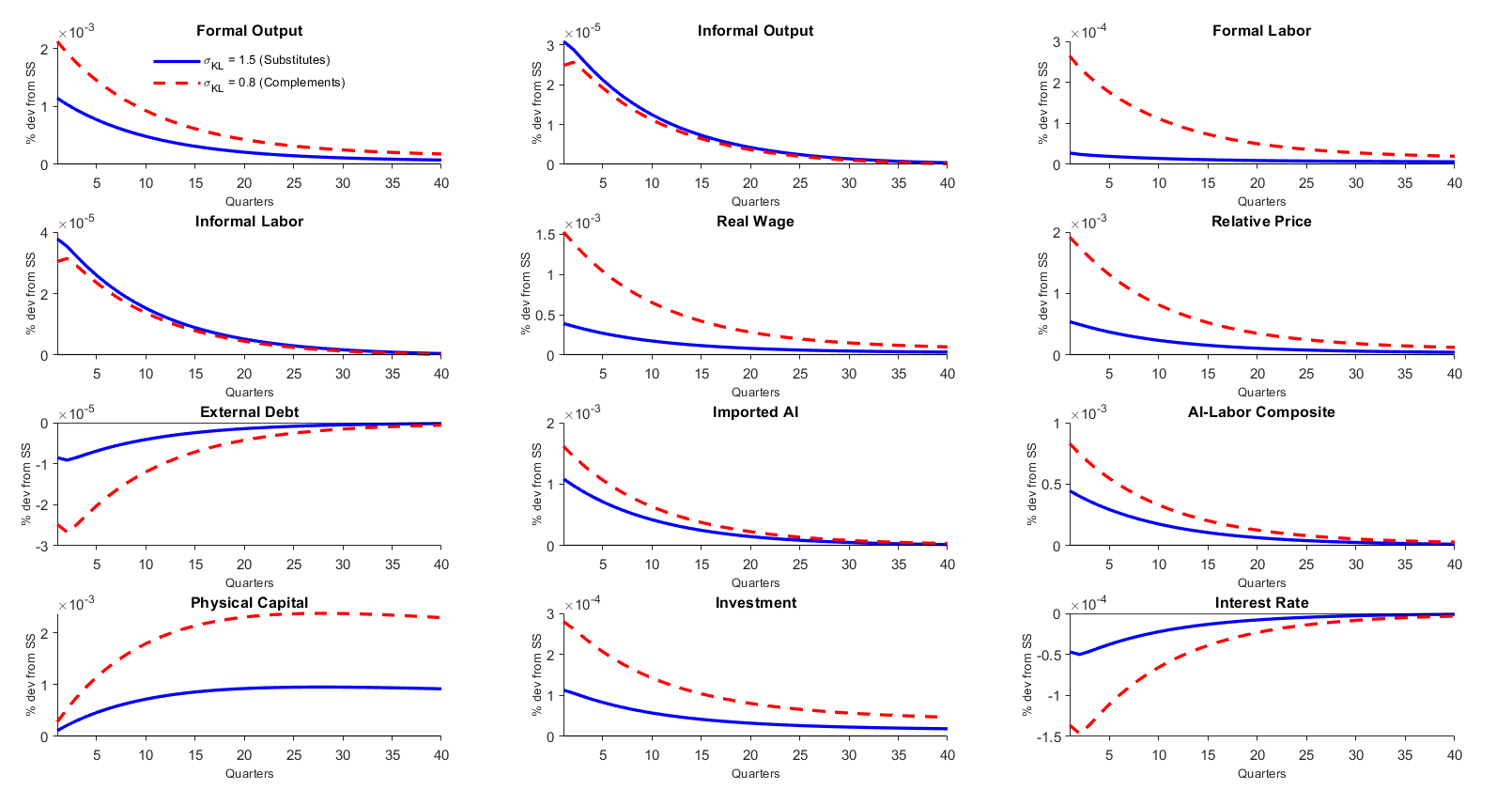}}
    \caption{Impulse responses to an international AI price shock under high consumption substitutability ($\eta = 1.5$). Blue solid lines represent substitution ($\sigma_{KL} = 1.5$) and red dashed lines represent complementarity ($\sigma_{KL} = 0.8$).}
    \label{fig:eta_150}
\end{figure}

This appendix explores the macroeconomic transmission of the international artificial intelligence price shock under an alternative demand-side regime where formal and informal goods are net substitutes in the household consumer basket, setting $\eta = 1.50$. Figure D1 presents the impulse response functions over forty quarters for both the substitution and complementarity technological specifications. The simulated trajectories show how high demand-side flexibility can amplify the expansionary effects of technology adoption and fundamentally alter labor reallocation patterns.When formal and informal commodities are easily substitutable in consumption, an increase in formal sector efficiency triggers a significant expenditure switch. The drop in imported technology costs lowers the marginal cost of formal firms, expanding formal output and driving down its relative price. Because consumers can readily shift their spending across sectors, they aggressively increase their consumption of formal goods. This demand reallocation creates a powerful product-market scale effect that dominates the production side of the economy. Most notably, as shown in the formal labor panel of Figure \ref{fig:eta_150}, formal employment expands on impact even under the technological substitution regime where $\sigma_{KL} = 1.5$. This result demonstrates that high consumption substitutability can completely override the partial equilibrium factor displacement effect inside the firm, turning an automation shock into a net generator of formal employment from the very first quarter.This adjustment is supported by the interaction between the real wage surge and the household labor supply rule. The aggregate efficiency boom caused by cheaper AI capital causes a substantial increase in the real wage, with the complementarity regime displaying a much larger amplitude. Under Greenwood, Hercowitz, and Huffman preferences, the negative wealth effect on leisure is neutralized, meaning that the aggregate supply of hours responds exclusively to the current real wage return. The sharp wage increase motivates households to expand the total labor force rapidly. This expansion of the aggregate labor pool increases total available hours fast enough to relax the sectoral trade-off. As a result, the formal sector expands its workforce without draining the informal buffer. Informal labor also increases to satisfy the non-tradable demand driven by higher aggregate income, which is reflected in the appreciation of the relative price of informal goods and the positive trajectory of informal output.The external sector and capital accumulation channels complete the general equilibrium feedback loop. The reduction in the international price of AI capital improves the trade balance, allowing the economy to reduce its external debt stock. Through the debt-elastic risk premium specification, this persistent deleveraging reduces country risk and drives down the domestic interest rate. The lower borrowing costs stimulate physical investment, which rises on impact and translates into a prominent, hump-shaped expansion of traditional physical capital. Because physical capital and the automated composite are combined via a unit-elastic Cobb-Douglas layer, this capital-deepening process continuously enhances the marginal productivity of workers over the medium run. This exercise reveals that when consumer demand is highly responsive, the benefits of technological change are maximized, allowing both sectors to expand while generating structural formalization.

\end{document}